\documentclass[aps,prl,twocolumn,superscriptaddress,longbibliography]{revtex4-1}
\usepackage{stix}
\usepackage{graphicx}
\usepackage{bm}
\usepackage{amsmath}
\usepackage{color}
\usepackage{color,amsmath}
\usepackage{graphicx,psfrag,accents,float}
\usepackage{multirow}
\usepackage{dcolumn}
\usepackage{xcolor}
\usepackage{comment}
\newcolumntype{L}[1]{>{\raggedright\arraybackslash}p{#1}}
\newcolumntype{C}[1]{>{\centering\arraybackslash}p{#1}}
\newcolumntype{R}[1]{>{\raggedleft\arraybackslash}p{#1}}

        \definecolor{AAcolor}{rgb}{0.7,0.1,0.4}

			\newcommand{\e}[1]{\begin{align}{#1}\end{align}}

		\newcommand{\f}[2]{\frac{#1}{#2}}

		\newcommand{\la}[1]{\label{#1}}

		\newcommand{\q}[1]{Eq.\ (\ref{#1})}

		\newcommand{\fig}[1]{Fig.\ \ref{#1}}

		\newcommand{\ri}{\rightarrow}

		\newcommand{\R}{\mathbb{R}}
		\newcommand{\Z}{\mathbb{Z}}

\newcommand{\var}{\varepsilon}

\newcommand\as{\;\;\;\;}

\newcommand{\ba}{\boldsymbol{a}}

\newcommand{\bk}{\boldsymbol{k}}

\newcommand{\bp}{\boldsymbol{p}}

\newcommand{\inv}{\mathfrak{i}}
\newcommand{\mir}{\mathfrak{r}}

\newcommand\rot{\mathfrak{c}}

\newcommand{\tz}{\tau_{\sma{3}}}

\newcommand{\invA}{\mathrm{\ang^{\text{-}\sma{1}}}}
\newcommand{\ang}{\mbox{\normalfont\AA}}
\newcommand{\invsqA}{\mathrm{\ang^{\text{-}\sma{2}}}}

\newcommand{\expect}[1]{\left\langle#1\right\rangle}

\newcommand{\lin}{\notag \\}

\newcommand{\bpm}{\begin{pmatrix}}
\newcommand{\epm}{\end{pmatrix}}

\newcommand{\bal}{\begin{align}}

\newcommand{\sma}[1]{\scriptscriptstyle{#1}}

\definecolor{CWcolor}{rgb}{0.1,0.5,0.4}

\begin{document}

\title{Diabolical touching point in the magnetic energy levels of topological nodal-line metals}

\author{Chong Wang}\affiliation{Department of Physics, Carnegie Mellon University, Pittsburgh, Pennsylvania 15213, USA}
\author{Zhongyi Zhang}\affiliation{Institute of Physics, Chinese Academy of Sciences, Beijing 100080, China}
\author{Chen Fang}\email{cfang@iphy.ac.cn}\affiliation{Institute of Physics, Chinese Academy of Sciences, Beijing 100080, China}
\author{A. Alexandradinata}\email{aalexan7@illinois.edu}\affiliation{Department of Physics and Institute for Condensed Matter Theory, University of Illinois at Urbana-Champaign, Urbana, Illinois 61801, USA}  

\begin{abstract}
For three-dimensional metals, Landau levels disperse as a function of the magnetic field and the momentum wavenumber parallel to the field. In this two-dimensional parameter space, it is shown that two conically-dispersing Landau levels can touch at a diabolical point -- a \emph{Landau-Dirac point}. The conditions giving rise to Landau-Dirac points are shown to be magnetic breakdown (field-induced quantum tunneling) and certain crystallographic spacetime symmetry. Both conditions are realizable in topological nodal-line metals, as we exemplify with CaP$_3$. A Landau-Dirac point reveals itself in anomalous ``batman''-like peaks in the magnetoresistance, as well as in the  onset of optical absorption linearly evolving to zero frequency as a function of the field magnitude/orientation. 
\end{abstract}
\date{\today}

\maketitle

For a real Hamiltonian, energy-level surfaces over a two-dimensional parameter space can locally form a double cone (\textit{diabolo}) with an energy-degenerate vertex known as a \textit{diabolical point}~\cite{neumann_wigner_eigenvalues,teller_crossingpotentialsurfaces,berry_semiclassicalmechanics,berry_diabolical}. The first physical application of the diabolical point was by W. R. Hamilton in his 1832 prediction of conical refraction~\cite{hamilton_conicaldiffraction,berry_conicaldiffraction}. {Since then, the diabolical point has re-emerged in diverse phenomena in singular optics~\cite{berry_singularoptics}, nuclear~\cite{herzberg_polyatomicdiabolicpoint,mead_bornoppenheimer,cederbaum_conicalintersections,farhan_diabolical} and quantum~\cite{ferretti_conicalintersection} physics. Its most recent revival is as Dirac-Weyl points~\cite{Nielsen_ABJanomaly_Weyl} in the crystal-momentum space of} topological semimetals~\cite{Horava_stabilityofFSandKtheory,Novoselov_graphene,wan_weylsemimetal,halasz_weylsemimetal,soluyanov_typeIIweyl} and insulators~\cite{kane2005A,fukanemele_3DTI,moore_3DTI,Rahul_3DTI}.



\begin{figure}
\centering
\includegraphics[width=1.0\columnwidth]{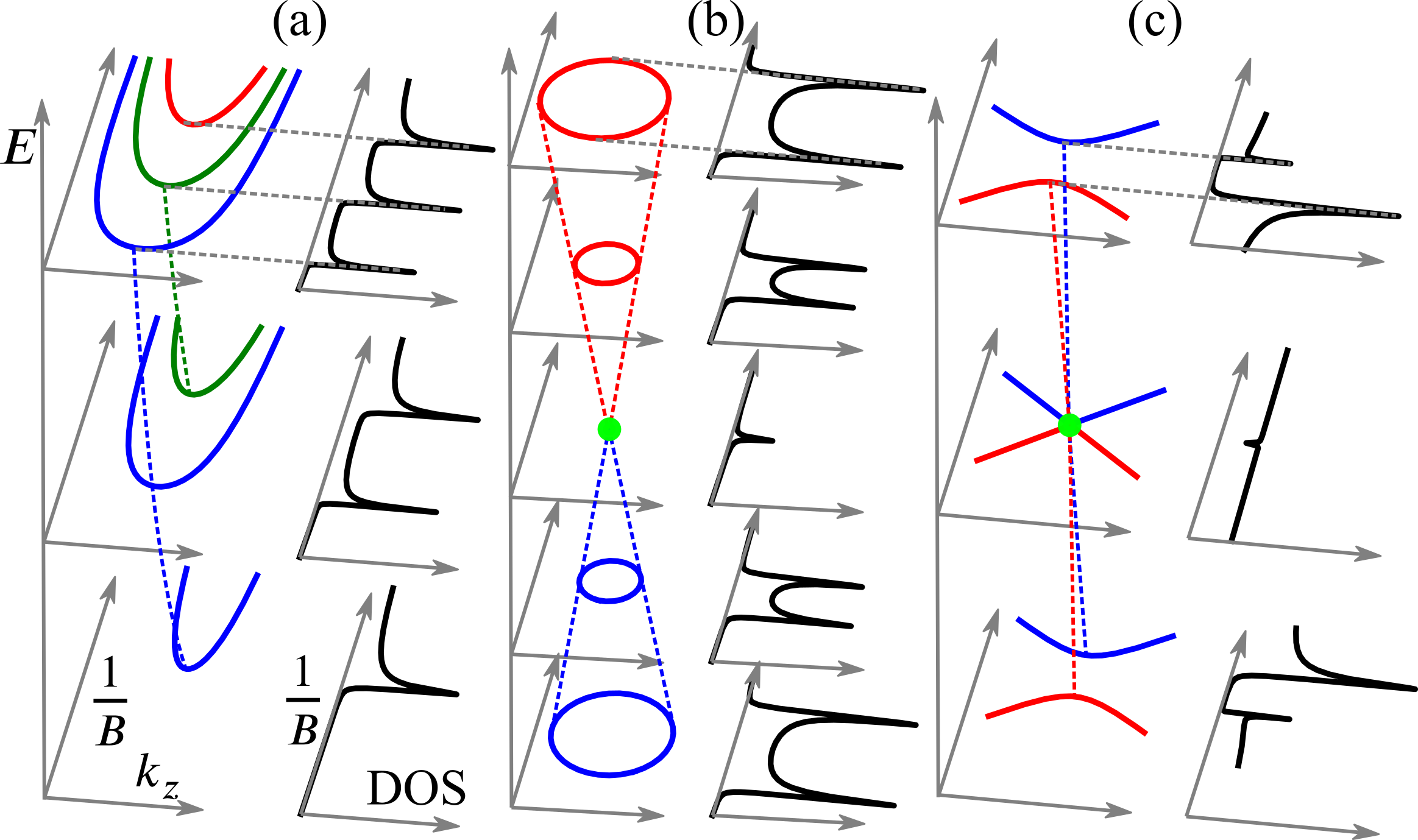}
\caption{Magnetic energy levels for a free-electron gas (a), and for topological nodal-line metals (b-c). Left of each panel: equienergy contours of energy-level surfaces in $(B^{\text{-}\sma{1}},k_z)$-space, with distinct sheets distinguished by color; right: corresponding density of states, regularized by a finite lifetime.}
\label{fig:cover_LandauDirac}
\end{figure}

{This work presents a heretofore undiscovered type of  diabolical point in a textbook solid-state phenomenon: the quantized energy spectrum of three-dimensional metals subject to a homogeneous magnetic field. A fundamental feature of the magnetic energy spectrum is its quantization into Landau levels~\cite{landaulifshitz_quantummechanics}, which are naturally parametrized by the field ($B$) and the momentum wavenumber ($k_z$) parallel to the field. In this two-dimensional parameter space,  \fig{fig:cover_LandauDirac}(b) illustrates how two Landau-level surfaces can touch at a diabolic point, which in the magnetic context will be referred to as a \textit{Landau-Dirac point}.
Parallel transport around an equienergy contour of the Landau-Dirac cone gives a  $\pi$ Berry phase~\cite{berry_quantalphase} which is topologically quantized.}

{Landau-Dirac points do \textit{not} exist for the free electron gas, nor do they exist for conventional metals with parabolic energy bands.
To appreciate this, consider that the Landau levels (in both cases)} are determined by the  Onsager-Lifshitz-Roth quantization rule~\cite{onsager,lifshitz_kosevich,rothII}: $\hbar/e B {=}{(2\pi n +\gamma)}/{S(E,k_z)},$ which is universally valid for weak fields.  $S(E,k_z)$ is the $\bk$-area enclosed by the orbit, $0{\leq }n{\in}\Z$ {is the Landau-level index}, and $\gamma$ is a subleading-in-$B$ correction inclusive of the Maslov~\cite{keller_correctedbohrsommerfeld} and Berry phases~\cite{rothII,mikitik_berryinmetal}, and a dynamical phase originating from the generalized Zeeman interaction~\cite{rothII,chang_niu_hyperorbit}. Henceforth, we set $\hbar {=} e {=} 1$ so that $B^{\text{-}\sma{1}}$ equals the square of the magnetic length.  {Generally for an electron-like (resp.\ hole-like) pocket, $S(E,k_z)$ is a single-valued function of $k_z$ and an increasing (resp.\ decreasing) function of energy $E$, e.g., $S{=}\pi(2mE-k_z^2)$ for a free-electron gas with mass $m$. These conditions on $S(E,k_z)$ ensure that} equienergy solutions of the quantization rule lie on \textit{open}, non-intersecting contours in $(B^{\text{-}\sma{1}},k_z)$-space, as illustrated for the free-electron gas in \fig{fig:cover_LandauDirac}(a). {It follows that the \textit{closed} equienergy contours  of the diabolo [cf.\ \fig{fig:cover_LandauDirac}(b)]  cannot derive from a single electron-like or hole-like pocket.}

However, if multiple pockets are linked by {field-driven quantum tunneling (known as magnetic breakdown~\cite{cohen_falicov_breakdown,azbel_quasiclassical,blount_effham,pippard1,chambers_breakdown})}, we will show that tunneling-induced level repulsion can convert open contours to  closed contours of a diabolo. {The stability of the Landau-Dirac point relies on} a certain crystallographic spacetime symmetry that is preserved in the presence of the field. For example, the composition $T\rot_{2y}$ of time reversal and two-fold rotation (about a field-orthogonal axis) maps $(B^{\text{-}\sma{1}},k_z){\ri} (B^{\text{-}\sma{1}},k_z)$, ensuring that Landau-Dirac points are movable over  $(B^{\text{-}\sma{1}},k_z)$-space, but irremovable unless annihilated in pairs -- as analogous to the Dirac points of graphene~\cite{Novoselov_graphene}. Either of spatial inversion $\inv$ [$(x,y,z){\ri}({-}x,{-}y,{-}z)]$ or reflection $\mir_z$ [$(x,y,z){\ri} (x,y,{-}z)]$ maps  $(B^{\text{-}\sma{1}},k_z){\ri} (B^{\text{-}\sma{1}},{-}k_z)$, and therefore crossings between Landau levels of opposite $\inv$ (or $\mir_z$) representations are perturbatively robust on high-symmetry lines. {All three symmetries, as well as the condition of magnetic breakdown, are realizable in topological nodal-line metals~\cite{burkov_linenodesemimetal,chenyuanping_weylloop,bzdusek_nodalchain,chingkai_classifysemimetal,yangbohm_toposemimetal,chenfang_nodallinereview} -- as we will first demonstrate with a conceptually-simple, minimal model, and subsequently for the nodal-line metallic candidate CaP$_3$. We will  show further that a Landau-Dirac point reveals itself in anomalous ``batman''-like peaks in the density of states [cf.\ \fig{fig:cover_LandauDirac}(b)], which leavesan experimental fingerprint in the magnetoresistance as well as in optical absorption.}

\begin{figure}
\centering
\includegraphics[width=1\columnwidth]{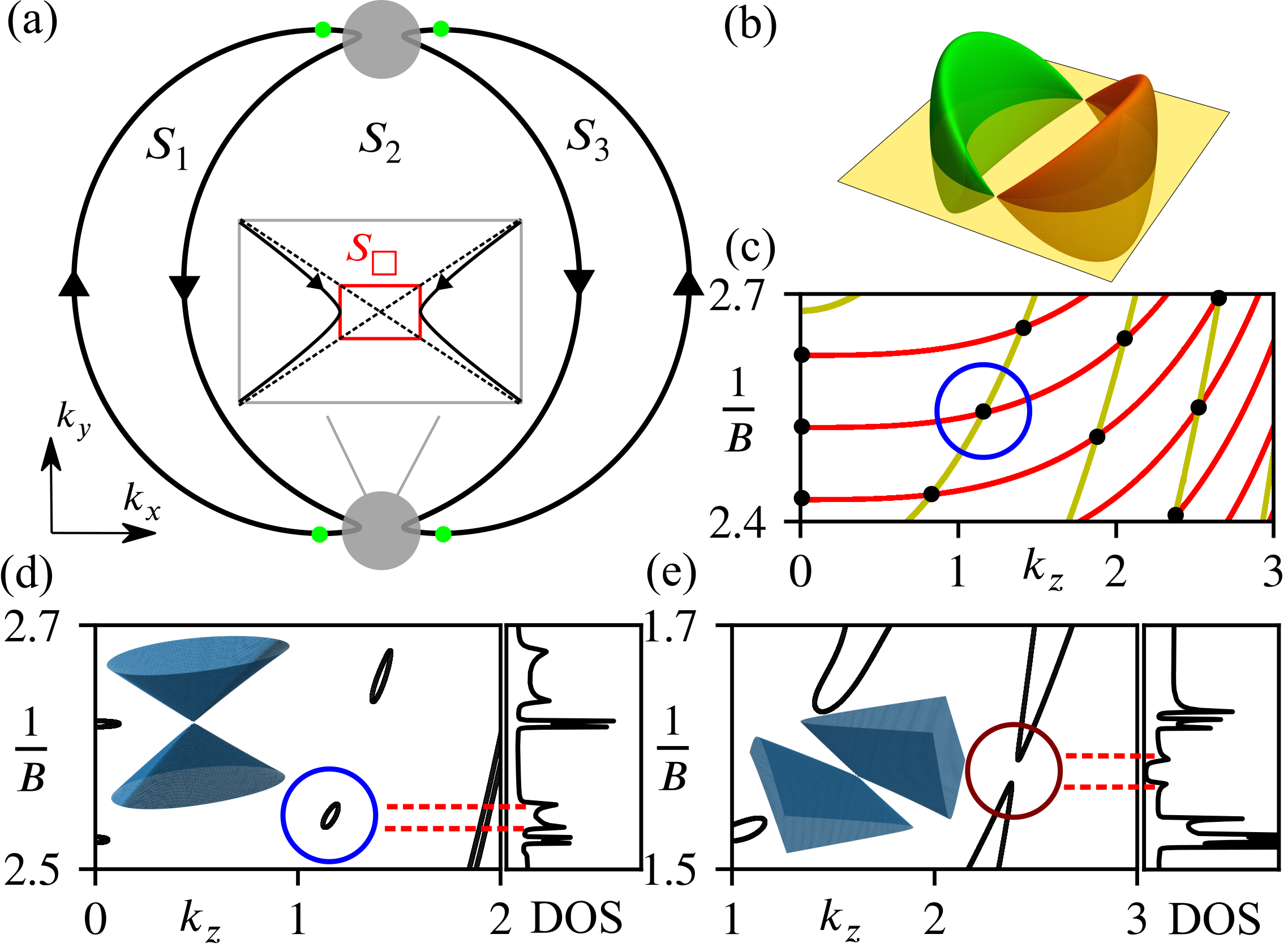}
\caption{ For the minimal model in \q{refsymm} with parameters: $v_x=v_z=m=1$ and $\epsilon_0=10$, we plot the zero-energy, Fermi surface within the Brillouin zone in panel (b); a constant-$(k_z{=}0.4)$ cross-section of the same surface is shown in (a). Inset of (a): enlarged view of breakdown region. Landau-Fermi surfaces over $(B^{\text{-}\sma{1}},k_z)$ are indicated by black dots and black lines in (c,d,e), for $E{=}0,0.01,0.95$ respectively. Right panels in (d,e) plot the corresponding DOS in arbitrary units. The diabolo in (d) [(e)] is the energy dispersion of the type-I (resp., type-II) Landau-Dirac cone encircled in blue (resp., brown). \label{fig:lldos}}
\end{figure}

\textit{Proof of principle.} We first present a minimal model of Landau-Dirac points with both $\mir_z$ and $T\rot_{2y}$ symmetries. At zero field, our effective-mass model describes two parabolic bands with opposite masses:
\e{ &H(\bk)= \left[({k_x^2+k_y^2})/{2m}-\var_0\right]\tau_3 + v_zk_z\tau_1+ v_xk_x. \la{refsymm}}
$\var_0$ being positive implies that the two bands overlap on the energy axis, however level repulsion is absent in the  $k_z{=}0$ plane owing to $\mir_z$ symmetry:  $\tau_3 H(\bk) \tau_3 {=} H(k_x,k_y,{-}k_z)$. It follows that a zero-energy, nodal-line degeneracy encircles $\bk{=}\bm{0}$ with radius $k_R{=}\sqrt{2m\var_0}$, supposing $v_x{=}0$. If nonzero, the $v_xk_x$ term causes the nodal line to disperse with bandwidth  $\Delta E{=}2v_xk_R$. Thus for a Fermi energy satisfying  $|E_F|{<}\Delta E/2$,  the Fermi surface comprises electron and hole pockets that interconnect like a linked sausage, as illustrated in  \fig{fig:lldos}(b). Close to either interconnection points (with $E_F{=}0$), an effective Hamiltonian  is attained by linearizing \q{refsymm} around $\bk{=}(0,\pm k_R,0)$:
\e{ H_{\pm} = \pm v_y \delta k_y\tau_3 + v_z k_z \tau_1+ v_xk_x, \as v_y=\sqrt{{2\var_0}/{m}}.\la{kdotp}}
whose equienergy contours form a hyperbola depicted in the inset of \fig{fig:lldos}(a). 

Applying a magnetic field parallel to ${-}z$, the magnetic energy levels are eigenvalues of the Peierls-Onsager Hamiltonian $H(K_x,K_y,k_z)$, which is obtained by the standard substitution of $(k_x,k_y)$  in the zero-field Hamiltonian [cf.\ \q{refsymm}] by non-commuting operators satisfying $[K_y,K_x]{=}iB$~\cite{peierls_substitution,luttinger_peierlssub}. If $B$ is much smaller than the $\bk$-area of both sausage-shaped pockets, the following semiclassical interpretation holds: the Lorentz force pushes electrons along semiclassical trajectories indicated by arrows in \fig{fig:lldos}(a). In the vicinity of both connection points [$\bk{=}(0,\pm k_R,0)$], inter-pocket tunneling occurs  with the Landau-Zener probability~\cite{blount_effham,zener_nonadiabaticcrossing}:
\e{\rho^2 = e^{-2\pi\mu}, \;\; \mu = {S_\square}/{8B}, \;\; S_\square={4v_z^2k_z^2}/{v_xv_y} ,\la{landauzenerprob}}
with $S_{\square}$ the rectangular area inscribed by the two hyperbolic arms [cf.\ inset of \fig{fig:lldos}(a)]. An analogous type of interband magnetic breakdown was first studied by  Slutskin~\cite{slutskin,kaganov_coherentmagneticbreakdown}, and developed by other authors in the context of Dirac-Weyl metals~\cite{obrien_breakdown,AALG_breakdown,100page,cwaa_landauquantization,breitkreiz_phaseshift}.
By matching the WKB wave functions~\cite{zilberman_wkb} at the tunneling regions (by the Landau-Zener connection formula~\cite{100page}), we arrive at the following quantization rule for the magnetic energy levels:
\e{ &0= Q(E,k_z,B^{\text{-}\sma{1}})=\cos X +\rho^2\cos Y +\tau^2\cos Z,\la{quantrule}\\
&(X,Y,Z)=\f{1}{2B}\big(S_1-S_3,S_{12}+S_{23}, S_{1}+S_{3}+4{\omega}B\big),\notag}
with $\tau^2{=}1{-}\rho^2$ the  probability that an incoming electron `reflects' off the tunneling region with a different velocity. $\omega {=} \mu{-}\mu \ln \mu {+}\text{arg}[\Gamma(i\mu)] {+}\pi/4$ is the phase acquired during this adiabatic reflection, with $\Gamma$ being the Gamma function; $S_1$ ($S_3$) is the $\bk$-area of the left (right) sausage-shaped pocket, and $S_{12}{:}{=}S_1{+}S_2$ ($S_{23}{:}{=}S_2{+}S_3$) is the area of the left (right) circular trajectory linked by tunneling [cf. Fig.~\ref{fig:lldos}(a)].

For $k_z{=}\mu{=}0$, Landau-Zener tunneling occurs with unit probability, and the solutions to \q{quantrule} describe independent cyclotron orbits over overlapping circles: 
\e{\cos X+\cos Y=0\;\Rightarrow \;  S_{12,23}(E,0)/B=2\pi(n+1/2). \la{onsagercircle}}
The zero-energy solutions of \q{onsagercircle} are doubly-degenerate and lie at equidistant points on the vertical axis of \fig{fig:lldos}(c), owing to the commensuration of areas: $S_{12}(0,k_z){=}S_{23}(0,k_z)$, which derives from the effective-mass Hamiltonian in \q{refsymm}.

There is no unique semiclassical trajectory in the  intermediate tunneling regime with nonzero but finite $\mu{\propto}k_z^2$.
We focus on a class of solutions contained in certain  hypersurfaces in $({E},{k}_{z},{B^{\text{-}\sma{1}}})$-space ($r$-space, in short), defined by $X(r)/\pi{\in}2\mathbb{Z}$ and $2\mathbb{Z}{+}1$. Whether even or odd, $\cos X$ is extremized to ${\pm} 1$, hence the \textit{only} solutions to \q{quantrule} must satisfy 
$\cos Y{=}\cos Z{=}{\mp}\cos X$. These being two constraints within a two-dimensional hypersurface, they can only be satisfied at isolated points which we denote by $\{\overline{r}\}$. Such points contained within the $(X{=}0)$ hypersurface are illustrated as black dots in \fig{fig:lldos}(c); note the $(X{=}0)$ hypersurface is just the $E{=}0$ plane owing to the just-mentioned commensuration condition, and the black dots lie at the intersections of red lines (defined by $\cos Y{=}{-}1$) and yellow lines ($\cos Z{=}{-}1$). 

As we move off a hypersurface in normal (or anti-normal) direction, each point solution evolves into an elliptical closed curve, as illustrated for $E{=}0.01$ in \fig{fig:lldos}(d). To demonstrate generally that $\overline{r}$ is a diabolical point, consider that $\overline{r}$ is an extremal point for each of $\{\cos X,\cos Y,\cos Z\}$. Consequently, for any solution of the quantization rule that deviates from $\overline{r}$ by small $\delta r{=}(\delta E,\delta k_z,\delta B^{\text{-}\sma{1}})$, $0{=} Q(\overline{r}+\delta r){-}Q(\overline{r})$ must be satisfied, with the right-hand side being quadratic in $\delta r$ to the lowest order. Solving this quadratic equation for the Landau-level dispersion,
\e{ &\delta E = \left({-b\pm \sqrt{b^2-4ac}}\right)/{2a}, \as a= X_{\sma{E}}^2-{\rho}^2 Y_{\sma{E}}^2-{\tau}^2 Z_{\sma{E}}^2, \lin
&b= \big[2X_{\sma{E}}(\delta k_z X_z+\delta B^{\text{-}\sma{1}} X_{B^{\sma{{-}1}}})\big] -{\rho}^2 [X{\ri}Y]-{\tau}^2 [X{\ri}Z], \lin
&c= \big[(\delta k_z X_z+\delta B^{\text{-}\sma{1}} X_{B^{\text{-}\sma{1}}})^2\big]-{\rho}^2 [X{\ri}Y]-{\tau}^2 [X{\ri}Z].\notag}
$X_{E,z,B^{\text{-}\sma{1}}}$ denotes the partial derivative of $X$ with respect to $(E,k_z,B^{\text{-}\sma{1}})$, as evaluated at $\overline{r}$; $[X{\ri}Y]$ denotes the substitution of $X$ with $Y$ in the square-bracketed expression on the same line. Since the quantity under the square root is quadratic in $(\delta k_z,\delta B^{\text{-}\sma{1}})$, the solution in $(\delta k_z,\delta B^{\text{-}\sma{1}})$-space generically forms a diabolo with vertex at $\overline{r}$. 

The perturbative stability of Landau-Dirac points is guaranteed by $T\rot_{2y}$ symmetry, which constrains the Peierls-Onsager Hamiltonian as $H(K_x,K_y,k_z)^*{=}H(K_x,{-}K_y,k_z)$. Given this anti-unitary constraint, a standard generalization~\cite{nogo_AAJH} of the von Neumann-Wigner theorem~\cite{neumann_wigner_eigenvalues} states that the codimension of an eigenvalue degeneracy is two, implying degeneracies are perturbatively stable in a two-dimensional parameter space -- given here by $(B^{\text{-}\sma{1}},k_z)$. The Landau-Dirac points at $k_z{=}0$ are doubly protected by $\mir_z$ symmetry, because each such point is a crossing between levels in distinct eigenspaces of $\tau_3$ (the matrix representation of $\mir_z$).


\textit{Type-II Landau-Dirac points.} While the $(X{=}0)$ hypersurface is simply the $E{=}0$ plane, $(X{=}\pi j)$ hypersurfaces are increasingly dispersive for larger $|j|$. With sufficient dispersion, the conical axis tilts so far from the energy axis, that the diabolo [centered at $(\bar{E},\bar{k}_z,\bar{B}^{\text{-}\sma{1}})$]  intersects the $E{=}\bar{E}$ plane on open lines; such a \textit{type-II Landau-Dirac point} occurs if and only if $ac{<}0$ on any segment of a circle encircling the diabolical point. A type-II point lying on the $X{=}6\pi$ hypersurface is illustrated in \fig{fig:lldos}(e).

An isolated, type-I point is distinguishable from type-II by the Fermi-level density of states (DOS). The intersection of a Landau-Dirac diabolo with the Fermi level defines a \textit{Landau-Fermi surface} parametrized by a multi-valued function $B^{\text{-}\sma{1}}(k_z)$ with two extrema, where the DOS diverges as two van-Hove singularities. The DOS in the vicinity of a single van-Hove peak is left-right asymmetric, being proportional to  $[\pm(B^{\text{-}\sma{1}}{-}B_0^{\text{-}\sma{1}})]^{\text{-}\sma{1/2}}$ on one side but not the other. (Such left-right asymmetry is routinely measured in thermodynamic/galvanomagnetic experiments on analogous solid-state systems~\cite{shoenberg_book,dhillon_bismuthmystery,guoAA_hightemperatureQO}.)
\fig{fig:cover_LandauDirac} illustrates that the inverse-square-root `tails' (in a type-I scenario) trail toward each other, resembling the helm of Batman; conversely, type-II tails trail apart, like anti-Batman.

\textit{Sum-over-histories approach to DOS.} The existence of (anti-)Batman peaks can be confirmed by computing the DOS from our quantization rule in \q{quantrule}. We offer an alternative  method of computation that is not only numerically efficient, but also instructively interprets the DOS -- as a sum of probability amplitudes for all possible closed-loop Feynman trajectories in $(k_x,k_y)$-space. Such trajectories are naturally described in the language of graphs~\cite{100page}: the equienergy contours of any band structure correspond to a graph with edges oriented by the Lorentz force; distinct edges are connected by two-in-two-out nodes, where interband tunneling or adiabatic reflection occurs. \fig{fig:lldos}(a) shows that the graph of our model has four edges (labelled $\alpha{=}1\ldots 4$) connected by two nodes (indicated by grey circles). 

A Feynman trajectory $\lambda$, defined as an ordered set of connected edges and nodes, is traversed with  probability amplitude $A_{\lambda}e^{i\phi_{\lambda}}$; $A_{\lambda}{\in}\R$ is a product of $\tau$ (one power for each reflection) and $\rho$ (one for each tunneling). The phase $\phi_{\lambda}$ sums contributions from edges and nodes: (a) an electron traversing an edge $\alpha{\in}\lambda$ [given by $k_x{=}k_x^{\alpha}(k_y)$] acquires a phase   $\varphi_{\alpha}{=}\int \text{-}k^{\alpha}_x dk_y/B{+}g_{\alpha}{+}m_{\alpha} \pi/2$, with the first term being the dynamical phase~\cite{onsager,lifshitz_kosevich}, the second the geometric Berry phase~\cite{rothII,mikitik_berryinmetal}, and the third the Maslov phase~\cite{keller_correctedbohrsommerfeld} from crossing  $m_{\alpha}$ number of turning points. (b)  An electron crossing a node acquires \textit{either} the phase $\omega$ (${-}\omega$) for adiabatic reflection within the higher-energy (resp. lower-energy) band, \textit{or} a $\pi$ phase~\cite{100page} for tunneling from lower- to higher-energy band, \textit{or} a trivial phase for tunneling from higher to lower energy. For example, the net phase acquired around the sausage loop (with area $S_1$) is $\phi_1{=}S_1/B{+}2\omega{+}\pi$, with $2\omega$ associated to two adiabatic reflections, and $\pi$ associated to two turning points indicated by {green dots} in \fig{fig:lldos}(a).

The DOS $\nu(\varepsilon)$, in units of the extensive Landau-level degeneracy $\mathcal{D}$, is expressible as a sum of amplitudes for all possible Feynman loops: \e{
\f{\nu}{\mathcal{D}}{=} \sum_{k_z}\bigg|\sum_{\alpha}\f{\partial_{\varepsilon} \varphi_{\alpha}}{2\pi}\big(1{+}2\,\text{Re}\,P_{\alpha\alpha}\big)\bigg|,\; P_{\beta\alpha}{=}\sum_{\lambda \in L_{\beta\alpha}} A_{\lambda} e^{i\phi_{\lambda}}.\la{dosform}
}
with $\text{Re}\,P{=}(P{+}P^*)/2$, and $L_{\alpha\beta}$ defined as the set of all trajectories emerging from the start point of  edge $\alpha$ and converging to the start point of edge $\beta$; $|\partial_{\varepsilon} \varphi_{\alpha}|$, to leading order in $B^{\text{-}\sma{1}}$, equals the time taken for an electron to traverse edge $\alpha$ following the semiclassical equation of motion. The above, formally-divergent formula is regularized by replacing $\varphi_{\beta}{\ri}\varphi_{\beta}{+}i0^+$ (for all $\beta$) in $P[\{\varphi_{\beta}\}]$. A similar formula for the DOS was first proposed by Kaganov and Slutskin~\cite{kaganov_coherentmagneticbreakdown} but contains a minor error that is clarified in the Supplemental Material~\cite{supplemental}.

To evaluate \q{dosform} efficiently, we exploit that $P_{\beta\alpha}$ satisfies a set of closed, recursive and linear equations, e.g., $P_{11}{=} \tau e^{i\omega} e^{i\phi_2} P_{21} {+} \rho e^{i\pi} e^{i\phi_3} P_{31}$
because all paths in $L_{11}$ must return to the start point of edge 1, either by reflection from edge 2 or by tunneling from edge 3. Each of $\{P_{\beta\alpha}\}_{\alpha,\beta{=}1\ldots 4}$ satisfies an analogous equation, giving an inhomogeneous system of 16 linear equations  with 16 unknowns, whose unique solution gives us the DOS via \q{dosform}. We  plot $\rho$ in the right panels of  \fig{fig:lldos}(d-e), with the correspondence between Batman peaks and type-I Landau-Fermi surfaces (anti-Batman and type-II) indicated by red dashed lines in \fig{fig:lldos}(d) [resp.\ \fig{fig:lldos}(e)]. 

Our sum-over-histories formula for the Batman peak manifests that it cannot be attributed to a unique semiclassical trajectory in the presence of intermediate tunneling strength. Consequently, Batman peaks are generally non-periodic in $1/B$ unlike conventional peaks in quantum oscillations; the width of the Batman helm is likewise \textit{not} attributable to the area of any $\bk$-loop in the graph. With multiple Landau-Dirac points, it is possible that batman and antibatman peaks overlap on the $B^{\text{-}\sma{1}}$ axis [as is nearly the case in \fig{fig:lldos}(d)], rendering their experimental identification ambiguous; this ambiguity is reduced by studying the evolution of the DOS as the candidate Landau-Dirac point is brought to the Fermi level, as will be explained near Letter-end.

\begin{figure}
\centering
\includegraphics[width=1\columnwidth]{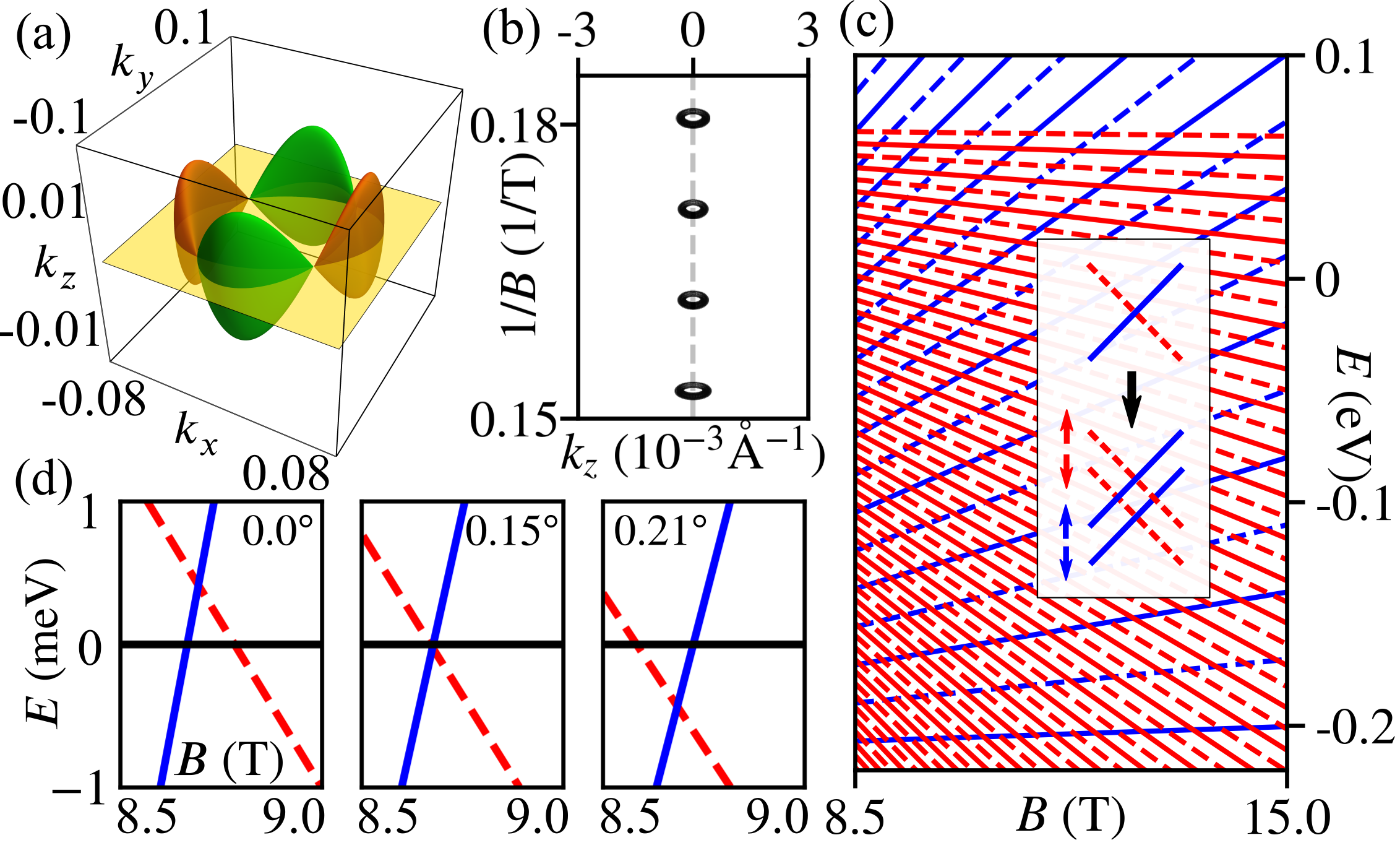}
\caption{For CaP$_3$ without spin-orbit coupling, we plot (a) the Fermi surface, (b)  Landau-Fermi surface, and (c) Landau-level dispersion at $k_z{=}0$ and field angle $\theta_B{=}0$. Panel (d) shows the dispersion of a specific Landau-Dirac point at $\theta_B{=}0^\circ,~0.15^\circ,~0.21^\circ$. Inset of (c) schematically illustrates a spin-split Landau-Dirac point.}\la{fig:cap3}
\end{figure}

\textit{CaP$_3$.} Our final case study is the time-reversal-invariant, nodal-line metal CaP$_3$, whose order-two point group is generated by spatial inversion $\inv$. Lacking fermiological studies of CaP$_3$, our case study is based on an ab-initio band-structure calculation by Xu et al~\cite{qiunan_cap3}. Among topological (semi)metal candidates, CaP$_3$, CaAs$_3$ and SrP$_3$ are unique  in having a Fermi surface that encloses a single, circular nodal line,  with no co-existing Fermi-surface pockets that are topologically trivial~\cite{qiunan_cap3,yquan_caas3}. CaP$_3$'s nodal line is centered at a inversion-invariant wavevector on the BZ boundary, and encircles an area ${\sim}1/50$ the areal dimension of the BZ -- this allows for an accurate description by an effective-mass Hamiltonian $H(\bk){=}\sum_{i=0}^3d_i(\bk)\tau_i$, with $\tau_0$ the identity matrix; the two-by-two matrix structure reflects our (present) ignorance of the weak spin-orbit interaction. 

Since the nodal line originates from an inversion of distinct representations of $\inv$, a basis may be found where $\inv$ constrains the Hamiltonian as $\tz H(\bk)\tz{=}H({-}\bk)$; the composition of time reversal and $\inv$ further constrains $\tz {H(\bk)}^*\tz{=}H(\bk)$, which implies $d_1{=}0$, so that nodal-line energy degeneracies are perturbatively stable by standard codimension arguments~\cite{neumann_wigner_eigenvalues}. Both symmetry constraints imply $d_0$ and $d_3$ are even functions of $\bk$, while $d_2{=}\ba{\cdot}\bk{+}O(k^3)$ with $\ba$ a real three-vector. It is convenient to perform an SO(3) $\bk$-rotation  so that $d_1{=}a_z  k_z$ (with $a_z{=}5.41$eV$\invA$) and $d_0$ is independent of $k_xk_y$, giving
    \begin{align}
        d_0=& {-}0.074{+} 54k_x^2 {+}5.2 k_y^2 {+}8.6k_z^2{-}1.8k_yk_z{+}11k_xk_z  \lin
        d_3=& {-}0.14 {+} 78k_x^2 {+}29 k_y^2 {+}45k_z^2{-}21k_xk_y{-}3.8k_yk_z{+}16k_xk_z \notag
    \end{align}
with all quadratic coefficients in units of eV$\invsqA$ and determined from ab-initio-fitted $\bk{\cdot}\bp$ parameters; the zero in energy is assumed to be the Fermi level, which we fix by  charge neutrality. The resultant Fermi surface consists of 
four pockets (two electron-like and two hole-like) which connect in the $k_z{=}0$ plane, as illustrated in \fig{fig:cap3}(a). $[H(k_x,k_y,k_z{=}0),\tz]{=}0$ reflects a $U(1){\times}U(1)$ symmetry of the effective-mass model which is absent in the lattice model. Note $H_{11}(k_x,k_y,0)$ is a parabolic dispersion with effective mass $m_{11}{=}0.12$ (in units of the free-electron mass); this parabola overlaps (on the energy axis) with the inverted parabola of $H_{22}$, with $m_{22}{=}0.31$. 


Applying a field in the ${-}z$ direction, both $\inv$ and $U(1){\times}U(1)$ symmetries are retained and constrain the Peierls-Onsager Hamiltonian as $\tz H(K_x,K_y,k_z)\tz{=}           H({-}K_x,{-}K_y,{-}k_z)$ and $[\tz,H(K_x,K_y,0)]{=}0$, respectively. $H_{11}(K_x,K_y,0)$ equivalently describes a quantum harmonic oscillator~\cite{landaulifshitz_quantummechanics}, while $H_{22}(K_x,K_y,0)$ describes an inverted oscillator with a cyclotron frequency that is smaller by the factor $m_{11}/m_{22}{=} 0.39$; both oscillator levels are plotted in \fig{fig:cap3}(c), with blue (red) lines indicating $\expect{\tau_z}{=}1$ ($\expect{\tau_z}{=}{-}1$). For either oscillator, the $\inv$ eigenvalue alternates between adjacent levels~\cite{landaulifshitz_quantummechanics}, as illustrated by alternating solid  ($\inv$-even) and dashed ($\inv$-odd) lines. Note that the zeroth level of $H_{11}$ (bottom-most blue line) is $\inv$-even, while that of  $H_{22}$ (top-most red line) is $\inv$-odd, owing to the distinct symmetry representations of basis vectors in the effective-mass Hamiltonian. Half of the Landau-Dirac points in \fig{fig:cap3}(c) are $\inv$-protected crossings between solid and dashed lines; the other half are protected by $U(1){\times}U(1)$ symmetry but not by $\inv$. The corresponding Landau-Fermi surface restricted to $k_z{=}0$ comprises a set of points [cf. \fig{fig:cap3}(b)].


For small $k_z{\neq}0$, the four sausage links in \fig{fig:cap3}(a) disconnect; electron dynamics in the vicinity of the four disconnected links  is again of the Landau-Zener type, with tunneling probability exp$(\text{-}2\pi \mu)$.   There is no unique, semiclassical trajectory \textit{except} at $k_z{=}0$, where tunneling occurs with unit probability. The previously-determined, point solutions (at $k_z{=}0$) extend horizontally to form closed lobes encircling  the type-I Landau-Dirac points at $k_z{=}0$, as shown in \fig{fig:cap3}(b). Just as for the previous, minimal model [cf.\ \fig{fig:lldos}(c-d)], we see the formation of closed Landau-Fermi surfaces in the $(B^{\text{-}\sma{1}},k_z)$-region where tunneling is intermediate in strength. 




Though our analysis has assumed a specific field orientation, half the crossings in \fig{fig:cap3}(c) are perturbatively stable against tilting of the field, because $\inv$ symmetry is maintained for any field orientation; the other half that relies on $U(1){\times}U(1)$ symmetry will destabilize. 

We have thus far neglected spin in the $CaP_3$ study. Accounting for the intrinsic Zeeman and spin-orbit interactions (which maintain $\inv$ symmetry), each spin-degenerate Landau level generically splits in energy, converting a single, spin-degenerate, $\inv$-protected Landau-Dirac point into four, spin-nondegenerate, $\inv$-protected Landau-Dirac points, as illustrated in the inset of \fig{fig:cap3}(c). An order-of-magnitude estimate for this energy splitting is given by the spin-orbit-induced splitting of the nodal-line degeneracy (at zero field),  which ranges from $4$ to $32$ meV~\cite{qiunan_cap3}.

Landau-Dirac points do not generically occur at the Fermi level. However, by tuning $B^{\text{-}\sma{1}}$, the two Landau levels (closest to the Fermi level) can be made to cross; by tuning a second parameter (e.g., the field orientation), such crossing, if $\inv$- or $\mir_z$-protected, can be brought to the Fermi level. \fig{fig:cap3}(d) illustrates a Landau-Dirac point being lowered to the Fermi level by tuning the tilt angle $\theta_{\sma{B}}$ in the $xz$ plane.
This manifests in the onset of optical absorption \textit{linearly} evolving to zero frequency as a function of $\theta_{\sma{B}}$ -- a smoking gun for the Landau-Dirac point.  Such an optical transition between Landau levels of distinct $\inv$ representations is allowed by the dipole selection rule~\cite{zawadzki_magnetoopticaltransitions}.
Simultaneously, the separation between a pair of (anti-)batman peaks in the DOS would linearly evolve to zero, as illustrated analogously in \fig{fig:cover_LandauDirac}(b-c). 

\textit{Outlook.} We began by presenting Landau-Dirac points in $(B^{\text{-}\sma{1}},k_z)$-space as the natural parameter space motivated by general symmetry considerations; on the other hand, Landau-Dirac points in $(B^{\text{-}\sma{1}},\theta_{\sma{B}})$-space are equally compelling for their experimental tunability. Topological nodal-line metals, including CaP$_3$, provide an experimental platform to make Landau-Dirac points a reality, owing to an interplay of magnetic symmetries and breakdown. Future investigations should determine if a similar Landau-Dirac phenomenology exists for {two other topological nodal-line material candidates, which host more complicated Fermi surfaces than the present study: (a)} SrAs$_3$, a cousin of CaP$_3$ with a similar crystalline structure and an experimentally-evidenced,\cite{shichao_sras3,linlin_sras3,hosen_sras3} nodal-line degeneracy, {and (b) the square-net compound ZrSiS, for which breakdown has been experimentally demonstrated~\cite{pezzini_breakdownZrSiS}.} Other platforms for Landau-Dirac points plausibly exist, owing to the diversity of magnetic space groups~\cite{bradley_magneticspacegroups}, as well as the qualitatively-distinct forms of breakdown in varied solid-state systems~\cite{cohen_falicov_breakdown,azbel_quasiclassical,blount_effham,pippard1,chambers_breakdown,slutskin,kaganov_coherentmagneticbreakdown,obrien_breakdown,AALG_breakdown,100page,cwaa_landauquantization,yuannoah_highordervanhove}. It is worth remarking that no symmetry is needed~\cite{neumann_wigner_eigenvalues} for stable Landau-Dirac points in a three-dimensional parameter space, e.g., $(B^{\text{-}\sma{1}},\theta_{\sma{B}},k_z)$ or $(B^{\text{-}\sma{1}},\theta_{\sma{B}},\theta'_{\sma{B}})$, with $\theta'_{\sma{B}}$ an independent tilt angle.

\begin{acknowledgments}
\textit{Acknowledgments.} We thank Di Xiao and Yang Gao for insightful discussions.  C.~W. was supported by the Department of Energy, Basic Energy Sciences, Materials Sciences and Engineering Division, Pro-QM EFRC (DE-SC0019443).
Z.~Z. and C.~F. were supported by the Ministry of Science and Technology of China under grant number 2016YFA0302400,
National Science Foundation of China under grant number 11674370, and
Chinese Academy of Sciences under grant numbers XXH13506-202 and XDB33000000. A.~A. was supported by the Gordon and Betty Moore Foundation EPiQS Initiative through Grant No. GBMF4305 at the University of Illinois. 

C.~W. and Z.~Z. contributed equally as co-first authors in this work.
\end{acknowledgments}

\bibliography{bib_Apr2018}

\begin{thebibliography}{70}%
\makeatletter
\providecommand \@ifxundefined [1]{%
 \@ifx{#1\undefined}
}%
\providecommand \@ifnum [1]{%
 \ifnum #1\expandafter \@firstoftwo
 \else \expandafter \@secondoftwo
 \fi
}%
\providecommand \@ifx [1]{%
 \ifx #1\expandafter \@firstoftwo
 \else \expandafter \@secondoftwo
 \fi
}%
\providecommand \natexlab [1]{#1}%
\providecommand \enquote  [1]{``#1''}%
\providecommand \bibnamefont  [1]{#1}%
\providecommand \bibfnamefont [1]{#1}%
\providecommand \citenamefont [1]{#1}%
\providecommand \href@noop [0]{\@secondoftwo}%
\providecommand \href [0]{\begingroup \@sanitize@url \@href}%
\providecommand \@href[1]{\@@startlink{#1}\@@href}%
\providecommand \@@href[1]{\endgroup#1\@@endlink}%
\providecommand \@sanitize@url [0]{\catcode `\\12\catcode `\$12\catcode
  `\&12\catcode `\#12\catcode `\^12\catcode `\_12\catcode `\%12\relax}%
\providecommand \@@startlink[1]{}%
\providecommand \@@endlink[0]{}%
\providecommand \url  [0]{\begingroup\@sanitize@url \@url }%
\providecommand \@url [1]{\endgroup\@href {#1}{\urlprefix }}%
\providecommand \urlprefix  [0]{URL }%
\providecommand \Eprint [0]{\href }%
\providecommand \doibase [0]{http://dx.doi.org/}%
\providecommand \selectlanguage [0]{\@gobble}%
\providecommand \bibinfo  [0]{\@secondoftwo}%
\providecommand \bibfield  [0]{\@secondoftwo}%
\providecommand \translation [1]{[#1]}%
\providecommand \BibitemOpen [0]{}%
\providecommand \bibitemStop [0]{}%
\providecommand \bibitemNoStop [0]{.\EOS\space}%
\providecommand \EOS [0]{\spacefactor3000\relax}%
\providecommand \BibitemShut  [1]{\csname bibitem#1\endcsname}%
\let\auto@bib@innerbib\@empty
\bibitem [{\citenamefont {von Neumann}\ and\ \citenamefont
  {Wigner}(1929)}]{neumann_wigner_eigenvalues}%
  \BibitemOpen
  \bibfield  {author} {\bibinfo {author} {\bibfnamefont {J.}~\bibnamefont {von
  Neumann}}\ and\ \bibinfo {author} {\bibfnamefont {E.}~\bibnamefont
  {Wigner}},\ }\bibfield  {title} {\enquote {\bibinfo {title} {On the behavior
  of eigenvalues in adiabatic processes},}\ }\href@noop {} {\bibfield
  {journal} {\bibinfo  {journal} {Phys. Z.}\ }\textbf {\bibinfo {volume}
  {30}},\ \bibinfo {pages} {467} (\bibinfo {year} {1929})}\BibitemShut
  {NoStop}%
\bibitem [{\citenamefont {Teller}(1937)}]{teller_crossingpotentialsurfaces}%
  \BibitemOpen
  \bibfield  {author} {\bibinfo {author} {\bibfnamefont {E.}~\bibnamefont
  {Teller}},\ }\bibfield  {title} {\enquote {\bibinfo {title} {The crossing of
  potential surfaces.}}\ }\href {\doibase 10.1021/j150379a010} {\bibfield
  {journal} {\bibinfo  {journal} {J. Phys. Chem.}\ }\textbf {\bibinfo {volume}
  {41}},\ \bibinfo {pages} {109--116} (\bibinfo {year} {1937})}\BibitemShut
  {NoStop}%
\bibitem [{\citenamefont {Berry}(1983)}]{berry_semiclassicalmechanics}%
  \BibitemOpen
  \bibfield  {author} {\bibinfo {author} {\bibfnamefont {Michael~Victor}\
  \bibnamefont {Berry}},\ }\bibfield  {title} {\enquote {\bibinfo {title}
  {Semiclassical mechanics of regular and irregular motion},}\ }in\ \href@noop
  {} {\emph {\bibinfo {booktitle} {Les Houches Lecture Series, vol. 36}}},\
  \bibinfo {editor} {edited by\ \bibinfo {editor} {\bibfnamefont {R.~H.
  G.~Helleman}\ \bibnamefont {G.~Iooss}}\ and\ \bibinfo {editor} {\bibfnamefont
  {R.}~\bibnamefont {Stora}}}\ (\bibinfo {address} {North-Holland, Amsterdam},\
  \bibinfo {year} {1983})\ p.\ \bibinfo {pages} {171–271}\BibitemShut
  {NoStop}%
\bibitem [{\citenamefont {Berry}\ and\ \citenamefont
  {Wilkinson}(1984)}]{berry_diabolical}%
  \BibitemOpen
  \bibfield  {author} {\bibinfo {author} {\bibfnamefont {Michael~Victor}\
  \bibnamefont {Berry}}\ and\ \bibinfo {author} {\bibfnamefont {Mark}\
  \bibnamefont {Wilkinson}},\ }\bibfield  {title} {\enquote {\bibinfo {title}
  {Diabolical points in the spectra of triangles},}\ }\href {\doibase
  10.1098/rspa.1984.0022} {\bibfield  {journal} {\bibinfo  {journal} {Proc. R.
  Soc. Lond. A}\ }\textbf {\bibinfo {volume} {392}},\ \bibinfo {pages} {15--43}
  (\bibinfo {year} {1984})}\BibitemShut {NoStop}%
\bibitem [{\citenamefont {Hamilton}(1837)}]{hamilton_conicaldiffraction}%
  \BibitemOpen
  \bibfield  {author} {\bibinfo {author} {\bibfnamefont {W.R.}\ \bibnamefont
  {Hamilton}},\ }\bibfield  {title} {\enquote {\bibinfo {title} {Third
  supplement to an essay on the theory of systems of rays},}\ }\href@noop {}
  {\bibfield  {journal} {\bibinfo  {journal} {Trans. R. Irish Acad.}\ }\textbf
  {\bibinfo {volume} {17}},\ \bibinfo {pages} {1--144} (\bibinfo {year}
  {1837})}\BibitemShut {NoStop}%
\bibitem [{\citenamefont {Berry}\ and\ \citenamefont
  {Jeffrey}(2007)}]{berry_conicaldiffraction}%
  \BibitemOpen
  \bibfield  {author} {\bibinfo {author} {\bibfnamefont {M.V.}\ \bibnamefont
  {Berry}}\ and\ \bibinfo {author} {\bibfnamefont {M.R.}\ \bibnamefont
  {Jeffrey}},\ }\bibfield  {title} {\enquote {\bibinfo {title} {Conical
  diffraction: Hamilton's diabolical point at the heart of crystal optics},}\
  }in\ \href {\doibase https://doi.org/10.1016/S0079-6638(07)50002-8} {\emph
  {\bibinfo {booktitle} {Progress in Optics}}},\ Vol.~\bibinfo {volume} {50},\
  \bibinfo {editor} {edited by\ \bibinfo {editor} {\bibfnamefont
  {E.}~\bibnamefont {Wolf}}}\ (\bibinfo  {publisher} {Elsevier},\ \bibinfo
  {address} {Amsterdam, Netherlands},\ \bibinfo {year} {2007})\ pp.\ \bibinfo
  {pages} {13 -- 50}\BibitemShut {NoStop}%
\bibitem [{\citenamefont {Berry}\ and\ \citenamefont
  {Dennis}(2003)}]{berry_singularoptics}%
  \BibitemOpen
  \bibfield  {author} {\bibinfo {author} {\bibfnamefont {M.~V.}\ \bibnamefont
  {Berry}}\ and\ \bibinfo {author} {\bibfnamefont {M.~R.}\ \bibnamefont
  {Dennis}},\ }\bibfield  {title} {\enquote {\bibinfo {title} {The optical
  singularities of birefringent dichroic chiral crystals},}\ }\href {\doibase
  10.1098/rspa.2003.1155} {\bibfield  {journal} {\bibinfo  {journal} {Proc. R.
  Soc. Lond. A}\ }\textbf {\bibinfo {volume} {459}},\ \bibinfo {pages}
  {1261--1292} (\bibinfo {year} {2003})}\BibitemShut {NoStop}%
\bibitem [{\citenamefont {Herzberg}\ and\ \citenamefont
  {Longuet-Higgins}(1963)}]{herzberg_polyatomicdiabolicpoint}%
  \BibitemOpen
  \bibfield  {author} {\bibinfo {author} {\bibfnamefont {G.}~\bibnamefont
  {Herzberg}}\ and\ \bibinfo {author} {\bibfnamefont {H.~C.}\ \bibnamefont
  {Longuet-Higgins}},\ }\bibfield  {title} {\enquote {\bibinfo {title}
  {Intersection of potential energy surfaces in polyatomic molecules},}\ }\href
  {\doibase 10.1039/DF9633500077} {\bibfield  {journal} {\bibinfo  {journal}
  {Discuss. Faraday Soc.}\ }\textbf {\bibinfo {volume} {35}},\ \bibinfo {pages}
  {77--82} (\bibinfo {year} {1963})}\BibitemShut {NoStop}%
\bibitem [{\citenamefont {Mead}\ and\ \citenamefont
  {Truhlar}(1979)}]{mead_bornoppenheimer}%
  \BibitemOpen
  \bibfield  {author} {\bibinfo {author} {\bibfnamefont {C.~Alden}\
  \bibnamefont {Mead}}\ and\ \bibinfo {author} {\bibfnamefont {Donald~G.}\
  \bibnamefont {Truhlar}},\ }\bibfield  {title} {\enquote {\bibinfo {title} {On
  the determination of born–oppenheimer nuclear motion wave functions
  including complications due to conical intersections and identical nuclei},}\
  }\href {\doibase 10.1063/1.437734} {\bibfield  {journal} {\bibinfo  {journal}
  {J. Chem. Phys.}\ }\textbf {\bibinfo {volume} {70}},\ \bibinfo {pages}
  {2284--2296} (\bibinfo {year} {1979})}\BibitemShut {NoStop}%
\bibitem [{\citenamefont {Cederbaum}\ \emph {et~al.}(2003)\citenamefont
  {Cederbaum}, \citenamefont {Friedman}, \citenamefont {Ryaboy},\ and\
  \citenamefont {Moiseyev}}]{cederbaum_conicalintersections}%
  \BibitemOpen
  \bibfield  {author} {\bibinfo {author} {\bibfnamefont {Lorenz~S.}\
  \bibnamefont {Cederbaum}}, \bibinfo {author} {\bibfnamefont {Ronald~S.}\
  \bibnamefont {Friedman}}, \bibinfo {author} {\bibfnamefont {Victor~M.}\
  \bibnamefont {Ryaboy}}, \ and\ \bibinfo {author} {\bibfnamefont {Nimrod}\
  \bibnamefont {Moiseyev}},\ }\bibfield  {title} {\enquote {\bibinfo {title}
  {Conical intersections and bound molecular states embedded in the
  continuum},}\ }\href {\doibase 10.1103/PhysRevLett.90.013001} {\bibfield
  {journal} {\bibinfo  {journal} {Phys. Rev. Lett.}\ }\textbf {\bibinfo
  {volume} {90}},\ \bibinfo {pages} {013001} (\bibinfo {year}
  {2003})}\BibitemShut {NoStop}%
\bibitem [{\citenamefont {Farhan}\ \emph {et~al.}(1996)\citenamefont {Farhan},
  \citenamefont {Canto}, \citenamefont {Rasmussen},\ and\ \citenamefont
  {Ring}}]{farhan_diabolical}%
  \BibitemOpen
  \bibfield  {author} {\bibinfo {author} {\bibfnamefont {A.R.}\ \bibnamefont
  {Farhan}}, \bibinfo {author} {\bibfnamefont {L.F.}\ \bibnamefont {Canto}},
  \bibinfo {author} {\bibfnamefont {J.O.}\ \bibnamefont {Rasmussen}}, \ and\
  \bibinfo {author} {\bibfnamefont {P.}~\bibnamefont {Ring}},\ }\bibfield
  {title} {\enquote {\bibinfo {title} {Form factors for two-nucleon transfer in
  the diabolical region of rotating nuclei},}\ }\href {\doibase
  https://doi.org/10.1016/0375-9474(95)00456-4} {\bibfield  {journal} {\bibinfo
   {journal} {Nucl. Phys. A}\ }\textbf {\bibinfo {volume} {597}},\ \bibinfo
  {pages} {387 -- 407} (\bibinfo {year} {1996})}\BibitemShut {NoStop}%
\bibitem [{\citenamefont {Ferretti}\ \emph {et~al.}(1999)\citenamefont
  {Ferretti}, \citenamefont {Lami},\ and\ \citenamefont
  {Villani}}]{ferretti_conicalintersection}%
  \BibitemOpen
  \bibfield  {author} {\bibinfo {author} {\bibfnamefont {Alessandro}\
  \bibnamefont {Ferretti}}, \bibinfo {author} {\bibfnamefont {Alessandro}\
  \bibnamefont {Lami}}, \ and\ \bibinfo {author} {\bibfnamefont {Giovanni}\
  \bibnamefont {Villani}},\ }\bibfield  {title} {\enquote {\bibinfo {title}
  {Transition probability due to a conical intersection: On the role of the
  initial conditions and of the geometric setup of the crossing surfaces},}\
  }\href {\doibase 10.1063/1.479375} {\bibfield  {journal} {\bibinfo  {journal}
  {The Journal of Chemical Physics}\ }\textbf {\bibinfo {volume} {111}},\
  \bibinfo {pages} {916--922} (\bibinfo {year} {1999})}\BibitemShut {NoStop}%
\bibitem [{\citenamefont {Nielsen}\ and\ \citenamefont
  {Ninomiya}(1983)}]{Nielsen_ABJanomaly_Weyl}%
  \BibitemOpen
  \bibfield  {author} {\bibinfo {author} {\bibfnamefont {H.B.}\ \bibnamefont
  {Nielsen}}\ and\ \bibinfo {author} {\bibfnamefont {Masao}\ \bibnamefont
  {Ninomiya}},\ }\bibfield  {title} {\enquote {\bibinfo {title} {The
  {Adler-Bell-Jackiw} anomaly and {Weyl} fermions in a crystal},}\ }\href
  {\doibase https://doi.org/10.1016/0370-2693(83)91529-0} {\bibfield  {journal}
  {\bibinfo  {journal} {Phys. Lett. B}\ }\textbf {\bibinfo {volume} {130}},\
  \bibinfo {pages} {389 -- 396} (\bibinfo {year} {1983})}\BibitemShut {NoStop}%
\bibitem [{\citenamefont {Ho\ifmmode~\check{r}\else
  \v{r}\fi{}ava}(2005)}]{Horava_stabilityofFSandKtheory}%
  \BibitemOpen
  \bibfield  {author} {\bibinfo {author} {\bibfnamefont {Petr}\ \bibnamefont
  {Ho\ifmmode~\check{r}\else \v{r}\fi{}ava}},\ }\bibfield  {title} {\enquote
  {\bibinfo {title} {Stability of fermi surfaces and {$K$} theory},}\ }\href
  {\doibase 10.1103/PhysRevLett.95.016405} {\bibfield  {journal} {\bibinfo
  {journal} {Phys. Rev. Lett.}\ }\textbf {\bibinfo {volume} {95}},\ \bibinfo
  {pages} {016405} (\bibinfo {year} {2005})}\BibitemShut {NoStop}%
\bibitem [{\citenamefont {Novoselov}\ \emph {et~al.}(2005)\citenamefont
  {Novoselov}, \citenamefont {Geim}, \citenamefont {Morozov}, \citenamefont
  {Jiang}, \citenamefont {Katsnelson}, \citenamefont {Grigorieva},
  \citenamefont {Dubonos},\ and\ \citenamefont {Firsov}}]{Novoselov_graphene}%
  \BibitemOpen
  \bibfield  {author} {\bibinfo {author} {\bibfnamefont {K.~S.}\ \bibnamefont
  {Novoselov}}, \bibinfo {author} {\bibfnamefont {A.~K.}\ \bibnamefont {Geim}},
  \bibinfo {author} {\bibfnamefont {S.~V.}\ \bibnamefont {Morozov}}, \bibinfo
  {author} {\bibfnamefont {D.}~\bibnamefont {Jiang}}, \bibinfo {author}
  {\bibfnamefont {M.~I.}\ \bibnamefont {Katsnelson}}, \bibinfo {author}
  {\bibfnamefont {I.~V.}\ \bibnamefont {Grigorieva}}, \bibinfo {author}
  {\bibfnamefont {S.~V.}\ \bibnamefont {Dubonos}}, \ and\ \bibinfo {author}
  {\bibfnamefont {A.~A.}\ \bibnamefont {Firsov}},\ }\bibfield  {title}
  {\enquote {\bibinfo {title} {Two-dimensional gas of massless {Dirac} fermions
  in graphene},}\ }\href {\doibase 10.1038/nature04233} {\bibfield  {journal}
  {\bibinfo  {journal} {Nature}\ }\textbf {\bibinfo {volume} {438}},\ \bibinfo
  {pages} {197--200} (\bibinfo {year} {2005})}\BibitemShut {NoStop}%
\bibitem [{\citenamefont {Wan}\ \emph {et~al.}(2011)\citenamefont {Wan},
  \citenamefont {Turner}, \citenamefont {Vishwanath},\ and\ \citenamefont
  {Savrasov}}]{wan_weylsemimetal}%
  \BibitemOpen
  \bibfield  {author} {\bibinfo {author} {\bibfnamefont {Xiangang}\
  \bibnamefont {Wan}}, \bibinfo {author} {\bibfnamefont {Ari}\ \bibnamefont
  {Turner}}, \bibinfo {author} {\bibfnamefont {Ashvin}\ \bibnamefont
  {Vishwanath}}, \ and\ \bibinfo {author} {\bibfnamefont {Sergey~Y.}\
  \bibnamefont {Savrasov}},\ }\bibfield  {title} {\enquote {\bibinfo {title}
  {Topological semimetal and {Fermi}-arc surface states in the electronic
  structure of pyrochlore iridates},}\ }\href@noop {} {\bibfield  {journal}
  {\bibinfo  {journal} {Phys. Rev. B}\ }\textbf {\bibinfo {volume} {83}},\
  \bibinfo {pages} {205101} (\bibinfo {year} {2011})}\BibitemShut {NoStop}%
\bibitem [{\citenamefont {Halasz}\ and\ \citenamefont
  {Balents}(2012)}]{halasz_weylsemimetal}%
  \BibitemOpen
  \bibfield  {author} {\bibinfo {author} {\bibfnamefont {Gabor~B.}\
  \bibnamefont {Halasz}}\ and\ \bibinfo {author} {\bibfnamefont {Leon}\
  \bibnamefont {Balents}},\ }\bibfield  {title} {\enquote {\bibinfo {title}
  {Time-reversal invariant realization of the weyl semimetal phase},}\
  }\href@noop {} {\bibfield  {journal} {\bibinfo  {journal} {Phys. Rev. B}\
  }\textbf {\bibinfo {volume} {85}},\ \bibinfo {pages} {035103} (\bibinfo
  {year} {2012})}\BibitemShut {NoStop}%
\bibitem [{\citenamefont {Soluyanov}\ \emph {et~al.}(2015)\citenamefont
  {Soluyanov}, \citenamefont {Gresch}, \citenamefont {Wang}, \citenamefont
  {Wu}, \citenamefont {Troyer}, \citenamefont {Dai},\ and\ \citenamefont
  {Bernevig}}]{soluyanov_typeIIweyl}%
  \BibitemOpen
  \bibfield  {author} {\bibinfo {author} {\bibfnamefont {Alexey~A}\
  \bibnamefont {Soluyanov}}, \bibinfo {author} {\bibfnamefont {Dominik}\
  \bibnamefont {Gresch}}, \bibinfo {author} {\bibfnamefont {Zhijun}\
  \bibnamefont {Wang}}, \bibinfo {author} {\bibfnamefont {QuanSheng}\
  \bibnamefont {Wu}}, \bibinfo {author} {\bibfnamefont {Matthias}\ \bibnamefont
  {Troyer}}, \bibinfo {author} {\bibfnamefont {Xi}~\bibnamefont {Dai}}, \ and\
  \bibinfo {author} {\bibfnamefont {B~Andrei}\ \bibnamefont {Bernevig}},\
  }\bibfield  {title} {\enquote {\bibinfo {title} {Type-{II} weyl
  semimetals},}\ }\href@noop {} {\bibfield  {journal} {\bibinfo  {journal}
  {Nature}\ }\textbf {\bibinfo {volume} {527}},\ \bibinfo {pages} {495--498}
  (\bibinfo {year} {2015})}\BibitemShut {NoStop}%
\bibitem [{\citenamefont {\textrm{C. L. Kane}}\ and\ \citenamefont {\textrm{E.
  J. Mele}}(2005)}]{kane2005A}%
  \BibitemOpen
  \bibfield  {author} {\bibinfo {author} {\bibnamefont {\textrm{C. L. Kane}}}\
  and\ \bibinfo {author} {\bibnamefont {\textrm{E. J. Mele}}},\ }\bibfield
  {title} {\enquote {\bibinfo {title} {\textrm{Quantum spin Hall effect in
  graphene}},}\ }\href@noop {} {\bibfield  {journal} {\bibinfo  {journal}
  {Phys. Rev. Lett.}\ }\textbf {\bibinfo {volume} {95}},\ \bibinfo {pages}
  {226801} (\bibinfo {year} {2005})}\BibitemShut {NoStop}%
\bibitem [{\citenamefont {Fu}\ \emph {et~al.}(2007)\citenamefont {Fu},
  \citenamefont {Kane},\ and\ \citenamefont {Mele}}]{fukanemele_3DTI}%
  \BibitemOpen
  \bibfield  {author} {\bibinfo {author} {\bibfnamefont {Liang}\ \bibnamefont
  {Fu}}, \bibinfo {author} {\bibfnamefont {C.~L.}\ \bibnamefont {Kane}}, \ and\
  \bibinfo {author} {\bibfnamefont {E.~J.}\ \bibnamefont {Mele}},\ }\bibfield
  {title} {\enquote {\bibinfo {title} {Topological insulators in three
  dimensions},}\ }\href {\doibase 10.1103/PhysRevLett.98.106803} {\bibfield
  {journal} {\bibinfo  {journal} {Phys. Rev. Lett.}\ }\textbf {\bibinfo
  {volume} {98}},\ \bibinfo {eid} {106803} (\bibinfo {year}
  {2007})}\BibitemShut {NoStop}%
\bibitem [{\citenamefont {Moore}\ and\ \citenamefont
  {Balents}(2007)}]{moore_3DTI}%
  \BibitemOpen
  \bibfield  {author} {\bibinfo {author} {\bibfnamefont {J.~E.}\ \bibnamefont
  {Moore}}\ and\ \bibinfo {author} {\bibfnamefont {L.}~\bibnamefont
  {Balents}},\ }\bibfield  {title} {\enquote {\bibinfo {title} {Topological
  invariants of time-reversal-invariant band structures},}\ }\href {\doibase
  10.1103/PhysRevB.75.121306} {\bibfield  {journal} {\bibinfo  {journal} {Phys.
  Rev. B}\ }\textbf {\bibinfo {volume} {75}},\ \bibinfo {eid} {121306}
  (\bibinfo {year} {2007})}\BibitemShut {NoStop}%
\bibitem [{\citenamefont {Roy}(2009)}]{Rahul_3DTI}%
  \BibitemOpen
  \bibfield  {author} {\bibinfo {author} {\bibfnamefont {Rahul}\ \bibnamefont
  {Roy}},\ }\bibfield  {title} {\enquote {\bibinfo {title} {Topological phases
  and the quantum spin hall effect in three dimensions},}\ }\href {\doibase
  10.1103/PhysRevB.79.195322} {\bibfield  {journal} {\bibinfo  {journal} {Phys.
  Rev. B}\ }\textbf {\bibinfo {volume} {79}},\ \bibinfo {pages} {195322}
  (\bibinfo {year} {2009})}\BibitemShut {NoStop}%
\bibitem [{\citenamefont {Landau}\ and\ \citenamefont
  {Lifshitz}(2007)}]{landaulifshitz_quantummechanics}%
  \BibitemOpen
  \bibfield  {author} {\bibinfo {author} {\bibfnamefont {L.~D.}\ \bibnamefont
  {Landau}}\ and\ \bibinfo {author} {\bibfnamefont {E.~M.}\ \bibnamefont
  {Lifshitz}},\ }\href@noop {} {\emph {\bibinfo {title} {Quantum Mechanics}}}\
  (\bibinfo  {publisher} {Elsevier},\ \bibinfo {address} {Singapore},\ \bibinfo
  {year} {2007})\BibitemShut {NoStop}%
\bibitem [{\citenamefont {Berry}(1984)}]{berry_quantalphase}%
  \BibitemOpen
  \bibfield  {author} {\bibinfo {author} {\bibfnamefont {M.~V.}\ \bibnamefont
  {Berry}},\ }\bibfield  {title} {\enquote {\bibinfo {title} {Quantal phase
  factors accompanying adiabatic changes},}\ }\href@noop {} {\bibfield
  {journal} {\bibinfo  {journal} {Proc. R. Soc. Lond A}\ }\textbf {\bibinfo
  {volume} {392}},\ \bibinfo {pages} {45} (\bibinfo {year} {1984})}\BibitemShut
  {NoStop}%
\bibitem [{\citenamefont {Onsager}(1952)}]{onsager}%
  \BibitemOpen
  \bibfield  {author} {\bibinfo {author} {\bibfnamefont {L.}~\bibnamefont
  {Onsager}},\ }\bibfield  {title} {\enquote {\bibinfo {title} {Interpretation
  of the de {Haas}-van {Alphen} effect},}\ }\href {\doibase
  10.1080/14786440908521019} {\bibfield  {journal} {\bibinfo  {journal}
  {Philos. Mag.}\ }\textbf {\bibinfo {volume} {43}},\ \bibinfo {pages}
  {1006--1008} (\bibinfo {year} {1952})}\BibitemShut {NoStop}%
\bibitem [{\citenamefont {Lifshitz}\ and\ \citenamefont
  {Kosevich}(1954)}]{lifshitz_kosevich}%
  \BibitemOpen
  \bibfield  {author} {\bibinfo {author} {\bibfnamefont {L.~M.}\ \bibnamefont
  {Lifshitz}}\ and\ \bibinfo {author} {\bibfnamefont {A.M.}\ \bibnamefont
  {Kosevich}},\ }\bibfield  {title} {\enquote {\bibinfo {title} {On the theory
  of the de {Haas}–van {Alphen} effect for particles with an arbitrary
  dispersion law},}\ }\href@noop {} {\bibfield  {journal} {\bibinfo  {journal}
  {Dokl. Akad. Nauk SSSR}\ }\textbf {\bibinfo {volume} {96}},\ \bibinfo {pages}
  {963} (\bibinfo {year} {1954})}\BibitemShut {NoStop}%
\bibitem [{\citenamefont {Roth}(1966)}]{rothII}%
  \BibitemOpen
  \bibfield  {author} {\bibinfo {author} {\bibfnamefont {Laura~M.}\
  \bibnamefont {Roth}},\ }\bibfield  {title} {\enquote {\bibinfo {title}
  {Semiclassical theory of magnetic energy levels and magnetic susceptibility
  of {Bloch} electrons},}\ }\href {\doibase 10.1103/PhysRev.145.434} {\bibfield
   {journal} {\bibinfo  {journal} {Phys. Rev.}\ }\textbf {\bibinfo {volume}
  {145}},\ \bibinfo {pages} {434--448} (\bibinfo {year} {1966})}\BibitemShut
  {NoStop}%
\bibitem [{\citenamefont {Keller}(1958)}]{keller_correctedbohrsommerfeld}%
  \BibitemOpen
  \bibfield  {author} {\bibinfo {author} {\bibfnamefont {Joseph~B.}\
  \bibnamefont {Keller}},\ }\bibfield  {title} {\enquote {\bibinfo {title}
  {Corrected {Bohr-Sommerfeld} quantum conditions for nonseparable systems},}\
  }\href {\doibase http://dx.doi.org/10.1016/0003-4916(58)90032-0} {\bibfield
  {journal} {\bibinfo  {journal} {Ann. Phys. (N. Y.)}\ }\textbf {\bibinfo
  {volume} {4}},\ \bibinfo {pages} {180 -- 188} (\bibinfo {year}
  {1958})}\BibitemShut {NoStop}%
\bibitem [{\citenamefont {Mikitik}\ and\ \citenamefont
  {Sharlai}(1999)}]{mikitik_berryinmetal}%
  \BibitemOpen
  \bibfield  {author} {\bibinfo {author} {\bibfnamefont {G.~P.}\ \bibnamefont
  {Mikitik}}\ and\ \bibinfo {author} {\bibfnamefont {Yu.~V.}\ \bibnamefont
  {Sharlai}},\ }\bibfield  {title} {\enquote {\bibinfo {title} {Manifestation
  of {Berry}'s phase in metal physics},}\ }\href {\doibase
  10.1103/PhysRevLett.82.2147} {\bibfield  {journal} {\bibinfo  {journal}
  {Phys. Rev. Lett.}\ }\textbf {\bibinfo {volume} {82}},\ \bibinfo {pages}
  {2147--2150} (\bibinfo {year} {1999})}\BibitemShut {NoStop}%
\bibitem [{\citenamefont {Chang}\ and\ \citenamefont
  {Niu}(1996)}]{chang_niu_hyperorbit}%
  \BibitemOpen
  \bibfield  {author} {\bibinfo {author} {\bibfnamefont {Ming-Che}\
  \bibnamefont {Chang}}\ and\ \bibinfo {author} {\bibfnamefont {Qian}\
  \bibnamefont {Niu}},\ }\bibfield  {title} {\enquote {\bibinfo {title}
  {{Berry} phase, hyperorbits, and the {Hofstadter} spectrum: Semiclassical
  dynamics in magnetic {Bloch} bands},}\ }\href {\doibase
  10.1103/PhysRevB.53.7010} {\bibfield  {journal} {\bibinfo  {journal} {Phys.
  Rev. B}\ }\textbf {\bibinfo {volume} {53}},\ \bibinfo {pages} {7010--7023}
  (\bibinfo {year} {1996})}\BibitemShut {NoStop}%
\bibitem [{\citenamefont {Cohen}\ and\ \citenamefont
  {Falicov}(1961)}]{cohen_falicov_breakdown}%
  \BibitemOpen
  \bibfield  {author} {\bibinfo {author} {\bibfnamefont {Morrel~H.}\
  \bibnamefont {Cohen}}\ and\ \bibinfo {author} {\bibfnamefont {L.~M.}\
  \bibnamefont {Falicov}},\ }\bibfield  {title} {\enquote {\bibinfo {title}
  {Magnetic breakdown in crystals},}\ }\href {\doibase
  10.1103/PhysRevLett.7.231} {\bibfield  {journal} {\bibinfo  {journal} {Phys.
  Rev. Lett.}\ }\textbf {\bibinfo {volume} {7}},\ \bibinfo {pages} {231--233}
  (\bibinfo {year} {1961})}\BibitemShut {NoStop}%
\bibitem [{\citenamefont {Azbel}(1961)}]{azbel_quasiclassical}%
  \BibitemOpen
  \bibfield  {author} {\bibinfo {author} {\bibfnamefont {M.~Ya.}\ \bibnamefont
  {Azbel}},\ }\bibfield  {title} {\enquote {\bibinfo {title} {Quasiclassical
  quantization in the neighborhood of singular classical trajectories},}\
  }\href@noop {} {\bibfield  {journal} {\bibinfo  {journal} {J. Exp. Theor.
  Phys.}\ }\textbf {\bibinfo {volume} {12}},\ \bibinfo {pages} {891} (\bibinfo
  {year} {1961})}\BibitemShut {NoStop}%
\bibitem [{\citenamefont {Blount}(1962)}]{blount_effham}%
  \BibitemOpen
  \bibfield  {author} {\bibinfo {author} {\bibfnamefont {E.~I.}\ \bibnamefont
  {Blount}},\ }\bibfield  {title} {\enquote {\bibinfo {title} {Bloch electrons
  in a magnetic field},}\ }\href {\doibase 10.1103/PhysRev.126.1636} {\bibfield
   {journal} {\bibinfo  {journal} {Phys. Rev.}\ }\textbf {\bibinfo {volume}
  {126}},\ \bibinfo {pages} {1636--1653} (\bibinfo {year} {1962})}\BibitemShut
  {NoStop}%
\bibitem [{\citenamefont {Pippard}(1962)}]{pippard1}%
  \BibitemOpen
  \bibfield  {author} {\bibinfo {author} {\bibfnamefont {A.~B.}\ \bibnamefont
  {Pippard}},\ }\bibfield  {title} {\enquote {\bibinfo {title} {Quantization of
  coupled orbits in metals},}\ }\href {\doibase 10.1098/rspa.1962.0200}
  {\bibfield  {journal} {\bibinfo  {journal} {Proc. R. Soc. Lond. A}\ }\textbf
  {\bibinfo {volume} {270}},\ \bibinfo {pages} {1--13} (\bibinfo {year}
  {1962})}\BibitemShut {NoStop}%
\bibitem [{\citenamefont {Chambers}(1966)}]{chambers_breakdown}%
  \BibitemOpen
  \bibfield  {author} {\bibinfo {author} {\bibfnamefont {W.~G.}\ \bibnamefont
  {Chambers}},\ }\bibfield  {title} {\enquote {\bibinfo {title} {Magnetic
  breakdown: Effective hamiltonian and de {Haas}-van {Alphen} effect},}\ }\href
  {\doibase 10.1103/PhysRev.149.493} {\bibfield  {journal} {\bibinfo  {journal}
  {Phys. Rev.}\ }\textbf {\bibinfo {volume} {149}},\ \bibinfo {pages}
  {493--504} (\bibinfo {year} {1966})}\BibitemShut {NoStop}%
\bibitem [{\citenamefont {Burkov}\ \emph {et~al.}(2011)\citenamefont {Burkov},
  \citenamefont {Hook},\ and\ \citenamefont
  {Balents}}]{burkov_linenodesemimetal}%
  \BibitemOpen
  \bibfield  {author} {\bibinfo {author} {\bibfnamefont {A.~A.}\ \bibnamefont
  {Burkov}}, \bibinfo {author} {\bibfnamefont {M.~D.}\ \bibnamefont {Hook}}, \
  and\ \bibinfo {author} {\bibfnamefont {Leon}\ \bibnamefont {Balents}},\
  }\bibfield  {title} {\enquote {\bibinfo {title} {Topological nodal
  semimetals},}\ }\href@noop {} {\bibfield  {journal} {\bibinfo  {journal}
  {Phys. Rev. B}\ }\textbf {\bibinfo {volume} {84}},\ \bibinfo {pages} {235126}
  (\bibinfo {year} {2011})}\BibitemShut {NoStop}%
\bibitem [{\citenamefont {Chen}\ \emph {et~al.}(2015)\citenamefont {Chen},
  \citenamefont {Xie}, \citenamefont {Yang}, \citenamefont {Pan}, \citenamefont
  {Zhang}, \citenamefont {Cohen},\ and\ \citenamefont
  {Zhang}}]{chenyuanping_weylloop}%
  \BibitemOpen
  \bibfield  {author} {\bibinfo {author} {\bibfnamefont {Yuanping}\
  \bibnamefont {Chen}}, \bibinfo {author} {\bibfnamefont {Yuee}\ \bibnamefont
  {Xie}}, \bibinfo {author} {\bibfnamefont {Shengyuan~A.}\ \bibnamefont
  {Yang}}, \bibinfo {author} {\bibfnamefont {Hui}\ \bibnamefont {Pan}},
  \bibinfo {author} {\bibfnamefont {Fan}\ \bibnamefont {Zhang}}, \bibinfo
  {author} {\bibfnamefont {Marvin~L.}\ \bibnamefont {Cohen}}, \ and\ \bibinfo
  {author} {\bibfnamefont {Shengbai}\ \bibnamefont {Zhang}},\ }\bibfield
  {title} {\enquote {\bibinfo {title} {Nanostructured carbon allotropes with
  {Weyl}-like loops and points},}\ }\href {\doibase
  10.1021/acs.nanolett.5b02978} {\bibfield  {journal} {\bibinfo  {journal}
  {Nano Lett.}\ }\textbf {\bibinfo {volume} {15}},\ \bibinfo {pages}
  {6974--6978} (\bibinfo {year} {2015})}\BibitemShut {NoStop}%
\bibitem [{\citenamefont {Bzdu{\v{s}}ek}\ \emph {et~al.}(2016)\citenamefont
  {Bzdu{\v{s}}ek}, \citenamefont {Wu}, \citenamefont {R{\"u}egg}, \citenamefont
  {Sigrist},\ and\ \citenamefont {Soluyanov}}]{bzdusek_nodalchain}%
  \BibitemOpen
  \bibfield  {author} {\bibinfo {author} {\bibfnamefont {Tom{\'a}{\v{s}}}\
  \bibnamefont {Bzdu{\v{s}}ek}}, \bibinfo {author} {\bibfnamefont {QuanSheng}\
  \bibnamefont {Wu}}, \bibinfo {author} {\bibfnamefont {Andreas}\ \bibnamefont
  {R{\"u}egg}}, \bibinfo {author} {\bibfnamefont {Manfred}\ \bibnamefont
  {Sigrist}}, \ and\ \bibinfo {author} {\bibfnamefont {Alexey~A.}\ \bibnamefont
  {Soluyanov}},\ }\bibfield  {title} {\enquote {\bibinfo {title} {Nodal-chain
  metals},}\ }\href {\doibase 10.1038/nature19099} {\bibfield  {journal}
  {\bibinfo  {journal} {Nature}\ }\textbf {\bibinfo {volume} {538}},\ \bibinfo
  {pages} {75--78} (\bibinfo {year} {2016})}\BibitemShut {NoStop}%
\bibitem [{\citenamefont {Chiu}\ and\ \citenamefont
  {Schnyder}(2014)}]{chingkai_classifysemimetal}%
  \BibitemOpen
  \bibfield  {author} {\bibinfo {author} {\bibfnamefont {Ching-Kai}\
  \bibnamefont {Chiu}}\ and\ \bibinfo {author} {\bibfnamefont {Andreas~P.}\
  \bibnamefont {Schnyder}},\ }\bibfield  {title} {\enquote {\bibinfo {title}
  {Classification of reflection-symmetry-protected topological semimetals and
  nodal superconductors},}\ }\href {\doibase 10.1103/PhysRevB.90.205136}
  {\bibfield  {journal} {\bibinfo  {journal} {Phys. Rev. B}\ }\textbf {\bibinfo
  {volume} {90}},\ \bibinfo {pages} {205136} (\bibinfo {year}
  {2014})}\BibitemShut {NoStop}%
\bibitem [{\citenamefont {Yang}\ \emph {et~al.}(2017)\citenamefont {Yang},
  \citenamefont {Bojesen}, \citenamefont {Morimoto},\ and\ \citenamefont
  {Furusaki}}]{yangbohm_toposemimetal}%
  \BibitemOpen
  \bibfield  {author} {\bibinfo {author} {\bibfnamefont {Bohm-Jung}\
  \bibnamefont {Yang}}, \bibinfo {author} {\bibfnamefont {Troels~Arnfred}\
  \bibnamefont {Bojesen}}, \bibinfo {author} {\bibfnamefont {Takahiro}\
  \bibnamefont {Morimoto}}, \ and\ \bibinfo {author} {\bibfnamefont {Akira}\
  \bibnamefont {Furusaki}},\ }\bibfield  {title} {\enquote {\bibinfo {title}
  {Topological semimetals protected by off-centered symmetries in nonsymmorphic
  crystals},}\ }\href {\doibase 10.1103/PhysRevB.95.075135} {\bibfield
  {journal} {\bibinfo  {journal} {Phys. Rev. B}\ }\textbf {\bibinfo {volume}
  {95}},\ \bibinfo {pages} {075135} (\bibinfo {year} {2017})}\BibitemShut
  {NoStop}%
\bibitem [{\citenamefont {Fang}\ \emph {et~al.}(2016)\citenamefont {Fang},
  \citenamefont {Weng}, \citenamefont {Dai},\ and\ \citenamefont
  {Fang}}]{chenfang_nodallinereview}%
  \BibitemOpen
  \bibfield  {author} {\bibinfo {author} {\bibfnamefont {Chen}\ \bibnamefont
  {Fang}}, \bibinfo {author} {\bibfnamefont {Hongming}\ \bibnamefont {Weng}},
  \bibinfo {author} {\bibfnamefont {Xi}~\bibnamefont {Dai}}, \ and\ \bibinfo
  {author} {\bibfnamefont {Zhong}\ \bibnamefont {Fang}},\ }\bibfield  {title}
  {\enquote {\bibinfo {title} {Topological nodal line semimetals},}\ }\href
  {\doibase 10.1088/1674-1056/25/11/117106} {\bibfield  {journal} {\bibinfo
  {journal} {Chin. Phys. B}\ }\textbf {\bibinfo {volume} {25}},\ \bibinfo
  {pages} {117106} (\bibinfo {year} {2016})}\BibitemShut {NoStop}%
\bibitem [{\citenamefont {Peierls}(1933)}]{peierls_substitution}%
  \BibitemOpen
  \bibfield  {author} {\bibinfo {author} {\bibfnamefont {R.}~\bibnamefont
  {Peierls}},\ }\bibfield  {title} {\enquote {\bibinfo {title} {{Zur} theorie
  des diamagnetismus von leitungselektronen},}\ }\href {\doibase
  10.1007/BF01342591} {\bibfield  {journal} {\bibinfo  {journal} {Z. Phys.}\
  }\textbf {\bibinfo {volume} {80}},\ \bibinfo {pages} {763--791} (\bibinfo
  {year} {1933})}\BibitemShut {NoStop}%
\bibitem [{\citenamefont {Luttinger}(1951)}]{luttinger_peierlssub}%
  \BibitemOpen
  \bibfield  {author} {\bibinfo {author} {\bibfnamefont {J.~M.}\ \bibnamefont
  {Luttinger}},\ }\bibfield  {title} {\enquote {\bibinfo {title} {The effect of
  a magnetic field on electrons in a periodic potential},}\ }\href {\doibase
  10.1103/PhysRev.84.814} {\bibfield  {journal} {\bibinfo  {journal} {Phys.
  Rev.}\ }\textbf {\bibinfo {volume} {84}},\ \bibinfo {pages} {814--817}
  (\bibinfo {year} {1951})}\BibitemShut {NoStop}%
\bibitem [{\citenamefont {Zener}\ and\ \citenamefont
  {Fowler}(1932)}]{zener_nonadiabaticcrossing}%
  \BibitemOpen
  \bibfield  {author} {\bibinfo {author} {\bibfnamefont {Clarence}\
  \bibnamefont {Zener}}\ and\ \bibinfo {author} {\bibfnamefont {Ralph~Howard}\
  \bibnamefont {Fowler}},\ }\bibfield  {title} {\enquote {\bibinfo {title}
  {Non-adiabatic crossing of energy levels},}\ }\href {\doibase
  10.1098/rspa.1932.0165} {\bibfield  {journal} {\bibinfo  {journal} {Proc. R.
  Soc. Lond. A}\ }\textbf {\bibinfo {volume} {137}},\ \bibinfo {pages}
  {696--702} (\bibinfo {year} {1932})}\BibitemShut {NoStop}%
\bibitem [{\citenamefont {Slutskin}(1968)}]{slutskin}%
  \BibitemOpen
  \bibfield  {author} {\bibinfo {author} {\bibfnamefont {A.A.}\ \bibnamefont
  {Slutskin}},\ }\bibfield  {title} {\enquote {\bibinfo {title} {Dynamics of
  conduction electrons under magnetic breakdown conditions},}\ }\href@noop {}
  {\bibfield  {journal} {\bibinfo  {journal} {J. Exp. Theor. Phys.}\ }\textbf
  {\bibinfo {volume} {26}},\ \bibinfo {pages} {474} (\bibinfo {year}
  {1968})}\BibitemShut {NoStop}%
\bibitem [{\citenamefont {Kaganov}\ and\ \citenamefont
  {Slutskin}(1983)}]{kaganov_coherentmagneticbreakdown}%
  \BibitemOpen
  \bibfield  {author} {\bibinfo {author} {\bibfnamefont {M.I.}\ \bibnamefont
  {Kaganov}}\ and\ \bibinfo {author} {\bibfnamefont {A.A.}\ \bibnamefont
  {Slutskin}},\ }\bibfield  {title} {\enquote {\bibinfo {title} {Coherent
  magnetic breakdown},}\ }\href {\doibase
  http://dx.doi.org/10.1016/0370-1573(83)90006-6} {\bibfield  {journal}
  {\bibinfo  {journal} {Phys. Rep.}\ }\textbf {\bibinfo {volume} {98}},\
  \bibinfo {pages} {189 -- 271} (\bibinfo {year} {1983})}\BibitemShut {NoStop}%
\bibitem [{\citenamefont {O'Brien}\ \emph {et~al.}(2016)\citenamefont
  {O'Brien}, \citenamefont {Diez},\ and\ \citenamefont
  {Beenakker}}]{obrien_breakdown}%
  \BibitemOpen
  \bibfield  {author} {\bibinfo {author} {\bibfnamefont {T.~E.}\ \bibnamefont
  {O'Brien}}, \bibinfo {author} {\bibfnamefont {M.}~\bibnamefont {Diez}}, \
  and\ \bibinfo {author} {\bibfnamefont {C.~W.~J.}\ \bibnamefont {Beenakker}},\
  }\bibfield  {title} {\enquote {\bibinfo {title} {Magnetic breakdown and
  {Klein} tunneling in a type-{II} {Weyl} semimetal},}\ }\href {\doibase
  10.1103/PhysRevLett.116.236401} {\bibfield  {journal} {\bibinfo  {journal}
  {Phys. Rev. Lett.}\ }\textbf {\bibinfo {volume} {116}},\ \bibinfo {pages}
  {236401} (\bibinfo {year} {2016})}\BibitemShut {NoStop}%
\bibitem [{\citenamefont {Alexandradinata}\ and\ \citenamefont
  {Glazman}(2017)}]{AALG_breakdown}%
  \BibitemOpen
  \bibfield  {author} {\bibinfo {author} {\bibfnamefont {A.}~\bibnamefont
  {Alexandradinata}}\ and\ \bibinfo {author} {\bibfnamefont {Leonid}\
  \bibnamefont {Glazman}},\ }\bibfield  {title} {\enquote {\bibinfo {title}
  {Geometric phase and orbital moment in quantization rules for magnetic
  breakdown},}\ }\href {\doibase 10.1103/PhysRevLett.119.256601} {\bibfield
  {journal} {\bibinfo  {journal} {Phys. Rev. Lett.}\ }\textbf {\bibinfo
  {volume} {119}},\ \bibinfo {pages} {256601} (\bibinfo {year}
  {2017})}\BibitemShut {NoStop}%
\bibitem [{\citenamefont {Alexandradinata}\ and\ \citenamefont
  {Glazman}(2018)}]{100page}%
  \BibitemOpen
  \bibfield  {author} {\bibinfo {author} {\bibfnamefont {A.}~\bibnamefont
  {Alexandradinata}}\ and\ \bibinfo {author} {\bibfnamefont {Leonid}\
  \bibnamefont {Glazman}},\ }\bibfield  {title} {\enquote {\bibinfo {title}
  {Semiclassical theory of landau levels and magnetic breakdown in topological
  metals},}\ }\href {\doibase 10.1103/PhysRevB.97.144422} {\bibfield  {journal}
  {\bibinfo  {journal} {Phys. Rev. B}\ }\textbf {\bibinfo {volume} {97}},\
  \bibinfo {pages} {144422} (\bibinfo {year} {2018})}\BibitemShut {NoStop}%
\bibitem [{\citenamefont {Wang}\ \emph {et~al.}(2019)\citenamefont {Wang},
  \citenamefont {Duan}, \citenamefont {Glazman},\ and\ \citenamefont
  {Alexandradinata}}]{cwaa_landauquantization}%
  \BibitemOpen
  \bibfield  {author} {\bibinfo {author} {\bibfnamefont {Chong}\ \bibnamefont
  {Wang}}, \bibinfo {author} {\bibfnamefont {Wenhui}\ \bibnamefont {Duan}},
  \bibinfo {author} {\bibfnamefont {Leonid}\ \bibnamefont {Glazman}}, \ and\
  \bibinfo {author} {\bibfnamefont {A.}~\bibnamefont {Alexandradinata}},\
  }\bibfield  {title} {\enquote {\bibinfo {title} {Landau quantization of
  nearly degenerate bands and full symmetry classification of landau level
  crossings},}\ }\href {\doibase 10.1103/PhysRevB.100.014442} {\bibfield
  {journal} {\bibinfo  {journal} {Phys. Rev. B}\ }\textbf {\bibinfo {volume}
  {100}},\ \bibinfo {pages} {014442} (\bibinfo {year} {2019})}\BibitemShut
  {NoStop}%
\bibitem [{\citenamefont {Breitkreiz}\ \emph {et~al.}(2018)\citenamefont
  {Breitkreiz}, \citenamefont {Bovenzi},\ and\ \citenamefont
  {Tworzyd\l{}o}}]{breitkreiz_phaseshift}%
  \BibitemOpen
  \bibfield  {author} {\bibinfo {author} {\bibfnamefont {M.}~\bibnamefont
  {Breitkreiz}}, \bibinfo {author} {\bibfnamefont {N.}~\bibnamefont {Bovenzi}},
  \ and\ \bibinfo {author} {\bibfnamefont {J.}~\bibnamefont {Tworzyd\l{}o}},\
  }\bibfield  {title} {\enquote {\bibinfo {title} {Phase shift of cyclotron
  orbits at type-{I} and type-{II} multi-{Weyl} nodes},}\ }\href {\doibase
  10.1103/PhysRevB.98.121403} {\bibfield  {journal} {\bibinfo  {journal} {Phys.
  Rev. B}\ }\textbf {\bibinfo {volume} {98}},\ \bibinfo {pages} {121403}
  (\bibinfo {year} {2018})}\BibitemShut {NoStop}%
\bibitem [{\citenamefont {Zil'berman}(1957)}]{zilberman_wkb}%
  \BibitemOpen
  \bibfield  {author} {\bibinfo {author} {\bibfnamefont {G.E.}\ \bibnamefont
  {Zil'berman}},\ }\bibfield  {title} {\enquote {\bibinfo {title} {Theory of
  {Bloch} electrons in a magnetic field},}\ }\href@noop {} {\bibfield
  {journal} {\bibinfo  {journal} {J. Exp. Theor. Phys.}\ }\textbf {\bibinfo
  {volume} {5}},\ \bibinfo {pages} {208} (\bibinfo {year} {1957})}\BibitemShut
  {NoStop}%
\bibitem [{\citenamefont {Alexandradinata}\ and\ \citenamefont
  {H\"oller}(2018)}]{nogo_AAJH}%
  \BibitemOpen
  \bibfield  {author} {\bibinfo {author} {\bibfnamefont {A.}~\bibnamefont
  {Alexandradinata}}\ and\ \bibinfo {author} {\bibfnamefont {J.}~\bibnamefont
  {H\"oller}},\ }\bibfield  {title} {\enquote {\bibinfo {title} {No-go theorem
  for topological insulators and high-throughput identification of {Chern}
  insulators},}\ }\href {\doibase 10.1103/PhysRevB.98.184305} {\bibfield
  {journal} {\bibinfo  {journal} {Phys. Rev. B}\ }\textbf {\bibinfo {volume}
  {98}},\ \bibinfo {pages} {184305} (\bibinfo {year} {2018})}\BibitemShut
  {NoStop}%
\bibitem [{\citenamefont {Shoenberg}(1984)}]{shoenberg_book}%
  \BibitemOpen
  \bibfield  {author} {\bibinfo {author} {\bibfnamefont {D.}~\bibnamefont
  {Shoenberg}},\ }\href@noop {} {\emph {\bibinfo {title} {Magnetic oscillations
  in metals}}}\ (\bibinfo  {publisher} {Cambridge University Press},\ \bibinfo
  {address} {The Edinburgh Building, Cambridge CB2 8RU, UK},\ \bibinfo {year}
  {1984})\BibitemShut {NoStop}%
\bibitem [{\citenamefont {Dhillon}\ and\ \citenamefont
  {Shoenberg}(1955)}]{dhillon_bismuthmystery}%
  \BibitemOpen
  \bibfield  {author} {\bibinfo {author} {\bibfnamefont {J.~S.}\ \bibnamefont
  {Dhillon}}\ and\ \bibinfo {author} {\bibfnamefont {David}\ \bibnamefont
  {Shoenberg}},\ }\bibfield  {title} {\enquote {\bibinfo {title} {The de
  {Haas}-van {Alphen} effect {III}. experiments at fields up to {32KG}},}\
  }\href {\doibase 10.1098/rsta.1955.0007} {\bibfield  {journal} {\bibinfo
  {journal} {Philos. Trans. Royal Soc. A}\ }\textbf {\bibinfo {volume} {248}},\
  \bibinfo {pages} {1--21} (\bibinfo {year} {1955})}\BibitemShut {NoStop}%
\bibitem [{\citenamefont {Guo}\ \emph {et~al.}(2019)\citenamefont {Guo},
  \citenamefont {Alexandradinata}, \citenamefont {Putzke}, \citenamefont {Fan},
  \citenamefont {Zhang}, \citenamefont {Wu}, \citenamefont {Yazyev},
  \citenamefont {Shirer}, \citenamefont {Bachmann}, \citenamefont {Bauer},
  \citenamefont {Ronning}, \citenamefont {Felser}, \citenamefont {Sun},\ and\
  \citenamefont {Moll}}]{guoAA_hightemperatureQO}%
  \BibitemOpen
  \bibfield  {author} {\bibinfo {author} {\bibfnamefont {Chunyu}\ \bibnamefont
  {Guo}}, \bibinfo {author} {\bibfnamefont {A.}~\bibnamefont
  {Alexandradinata}}, \bibinfo {author} {\bibfnamefont {Carsten}\ \bibnamefont
  {Putzke}}, \bibinfo {author} {\bibfnamefont {Feng-Ren}\ \bibnamefont {Fan}},
  \bibinfo {author} {\bibfnamefont {Shengnan}\ \bibnamefont {Zhang}}, \bibinfo
  {author} {\bibfnamefont {Quansheng}\ \bibnamefont {Wu}}, \bibinfo {author}
  {\bibfnamefont {Oleg~V.}\ \bibnamefont {Yazyev}}, \bibinfo {author}
  {\bibfnamefont {Kent~R.}\ \bibnamefont {Shirer}}, \bibinfo {author}
  {\bibfnamefont {Maja~D.}\ \bibnamefont {Bachmann}}, \bibinfo {author}
  {\bibfnamefont {Eric~D.}\ \bibnamefont {Bauer}}, \bibinfo {author}
  {\bibfnamefont {Filip}\ \bibnamefont {Ronning}}, \bibinfo {author}
  {\bibfnamefont {Claudia}\ \bibnamefont {Felser}}, \bibinfo {author}
  {\bibfnamefont {Yan}\ \bibnamefont {Sun}}, \ and\ \bibinfo {author}
  {\bibfnamefont {Philip J.~W.}\ \bibnamefont {Moll}},\ }\href@noop {}
  {\enquote {\bibinfo {title} {Fingerprint of topology in high-temperature
  quantum oscillations},}\ } (\bibinfo {year} {2019}),\ \Eprint
  {http://arxiv.org/abs/1910.07608} {arXiv:1910.07608 [cond-mat.str-el]}
  \BibitemShut {NoStop}%
\bibitem [{sup()}]{supplemental}%
  \BibitemOpen
  \href@noop {} {}\bibinfo {note} {Supplemental Material, which cites Refs.
  \onlinecite{onsager,lifshitz_kosevich_jetp,keller_correctedbohrsommerfeld,rothII,mikitik_berryinmetal,100page,zener_nonadiabaticcrossing,100page,falicov_dhva_magnesium,kaganov_coherentmagneticbreakdown,qiunan_cap3,cwaa_landauquantization,zak_magnetictranslation,Brown_magnetictranslation,landaulifshitz_quantummechanics}}\BibitemShut
  {NoStop}%
\bibitem [{\citenamefont {Xu}\ \emph {et~al.}(2017)\citenamefont {Xu},
  \citenamefont {Yu}, \citenamefont {Fang}, \citenamefont {Dai},\ and\
  \citenamefont {Weng}}]{qiunan_cap3}%
  \BibitemOpen
  \bibfield  {author} {\bibinfo {author} {\bibfnamefont {Qiunan}\ \bibnamefont
  {Xu}}, \bibinfo {author} {\bibfnamefont {Rui}\ \bibnamefont {Yu}}, \bibinfo
  {author} {\bibfnamefont {Zhong}\ \bibnamefont {Fang}}, \bibinfo {author}
  {\bibfnamefont {Xi}~\bibnamefont {Dai}}, \ and\ \bibinfo {author}
  {\bibfnamefont {Hongming}\ \bibnamefont {Weng}},\ }\bibfield  {title}
  {\enquote {\bibinfo {title} {Topological nodal line semimetals in the
  {CaP}$_{3}$ family of materials},}\ }\href {\doibase
  10.1103/PhysRevB.95.045136} {\bibfield  {journal} {\bibinfo  {journal} {Phys.
  Rev. B}\ }\textbf {\bibinfo {volume} {95}},\ \bibinfo {pages} {045136}
  (\bibinfo {year} {2017})}\BibitemShut {NoStop}%
\bibitem [{\citenamefont {Quan}\ \emph {et~al.}(2017)\citenamefont {Quan},
  \citenamefont {Yin},\ and\ \citenamefont {Pickett}}]{yquan_caas3}%
  \BibitemOpen
  \bibfield  {author} {\bibinfo {author} {\bibfnamefont {Y.}~\bibnamefont
  {Quan}}, \bibinfo {author} {\bibfnamefont {Z.~P.}\ \bibnamefont {Yin}}, \
  and\ \bibinfo {author} {\bibfnamefont {W.~E.}\ \bibnamefont {Pickett}},\
  }\bibfield  {title} {\enquote {\bibinfo {title} {Single nodal loop of
  accidental degeneracies in minimal symmetry: Triclinic {CaAs}$_3$},}\ }\href
  {\doibase 10.1103/PhysRevLett.118.176402} {\bibfield  {journal} {\bibinfo
  {journal} {Phys. Rev. Lett.}\ }\textbf {\bibinfo {volume} {118}},\ \bibinfo
  {pages} {176402} (\bibinfo {year} {2017})}\BibitemShut {NoStop}%
\bibitem [{\citenamefont
  {Zawadzki}(1991)}]{zawadzki_magnetoopticaltransitions}%
  \BibitemOpen
  \bibfield  {author} {\bibinfo {author} {\bibfnamefont {Wlodek}\ \bibnamefont
  {Zawadzki}},\ }\bibfield  {title} {\enquote {\bibinfo {title} {Intraband and
  interband magneto-optical transitions in semiconductors},}\ }in\ \href@noop
  {} {\emph {\bibinfo {booktitle} {Landau level spectroscopy}}},\ \bibinfo
  {editor} {edited by\ \bibinfo {editor} {\bibfnamefont {G.}~\bibnamefont
  {Landwehr}}\ and\ \bibinfo {editor} {\bibfnamefont {E.~I.}\ \bibnamefont
  {Rashba}}}\ (\bibinfo  {publisher} {Elsevier},\ \bibinfo {address}
  {North-Holland, Amsterdam},\ \bibinfo {year} {1991})\ p.\ \bibinfo {pages}
  {485}\BibitemShut {NoStop}%
\bibitem [{\citenamefont {Li}\ \emph {et~al.}(2018)\citenamefont {Li},
  \citenamefont {Guo}, \citenamefont {Fu}, \citenamefont {Pan}, \citenamefont
  {Wang}, \citenamefont {Ran}, \citenamefont {Bao}, \citenamefont {Ma},
  \citenamefont {Cai}, \citenamefont {Wang}, \citenamefont {Yu}, \citenamefont
  {Sun}, \citenamefont {Song},\ and\ \citenamefont {Wen}}]{shichao_sras3}%
  \BibitemOpen
  \bibfield  {author} {\bibinfo {author} {\bibfnamefont {Shichao}\ \bibnamefont
  {Li}}, \bibinfo {author} {\bibfnamefont {Zhaopeng}\ \bibnamefont {Guo}},
  \bibinfo {author} {\bibfnamefont {Dongzhi}\ \bibnamefont {Fu}}, \bibinfo
  {author} {\bibfnamefont {Xing-Chen}\ \bibnamefont {Pan}}, \bibinfo {author}
  {\bibfnamefont {Jinghui}\ \bibnamefont {Wang}}, \bibinfo {author}
  {\bibfnamefont {Kejing}\ \bibnamefont {Ran}}, \bibinfo {author}
  {\bibfnamefont {Song}\ \bibnamefont {Bao}}, \bibinfo {author} {\bibfnamefont
  {Zhen}\ \bibnamefont {Ma}}, \bibinfo {author} {\bibfnamefont {Zhengwei}\
  \bibnamefont {Cai}}, \bibinfo {author} {\bibfnamefont {Rui}\ \bibnamefont
  {Wang}}, \bibinfo {author} {\bibfnamefont {Rui}\ \bibnamefont {Yu}}, \bibinfo
  {author} {\bibfnamefont {Jian}\ \bibnamefont {Sun}}, \bibinfo {author}
  {\bibfnamefont {Fengqi}\ \bibnamefont {Song}}, \ and\ \bibinfo {author}
  {\bibfnamefont {Jinsheng}\ \bibnamefont {Wen}},\ }\bibfield  {title}
  {\enquote {\bibinfo {title} {Evidence for a {Dirac} nodal-line semimetal in
  {SrAs}$_3$},}\ }\href {\doibase https://doi.org/10.1016/j.scib.2018.04 .011}
  {\bibfield  {journal} {\bibinfo  {journal} {Sci. Bull.}\ }\textbf {\bibinfo
  {volume} {63}},\ \bibinfo {pages} {535 -- 541} (\bibinfo {year}
  {2018})}\BibitemShut {NoStop}%
\bibitem [{\citenamefont {An}\ \emph {et~al.}(2019)\citenamefont {An},
  \citenamefont {Zhu}, \citenamefont {Gao}, \citenamefont {Wu}, \citenamefont
  {Ning},\ and\ \citenamefont {Tian}}]{linlin_sras3}%
  \BibitemOpen
  \bibfield  {author} {\bibinfo {author} {\bibfnamefont {Linlin}\ \bibnamefont
  {An}}, \bibinfo {author} {\bibfnamefont {Xiangde}\ \bibnamefont {Zhu}},
  \bibinfo {author} {\bibfnamefont {Wenshuai}\ \bibnamefont {Gao}}, \bibinfo
  {author} {\bibfnamefont {Min}\ \bibnamefont {Wu}}, \bibinfo {author}
  {\bibfnamefont {Wei}\ \bibnamefont {Ning}}, \ and\ \bibinfo {author}
  {\bibfnamefont {Mingliang}\ \bibnamefont {Tian}},\ }\bibfield  {title}
  {\enquote {\bibinfo {title} {Chiral anomaly and nontrivial {Berry} phase in
  the topological nodal-line semimetal $\mathrm{SrA}{\mathrm{s}}_{3}$},}\
  }\href {\doibase 10.1103/PhysRevB.99.045143} {\bibfield  {journal} {\bibinfo
  {journal} {Phys. Rev. B}\ }\textbf {\bibinfo {volume} {99}},\ \bibinfo
  {pages} {045143} (\bibinfo {year} {2019})}\BibitemShut {NoStop}%
\bibitem [{\citenamefont {Hosen}\ \emph {et~al.}(2020)\citenamefont {Hosen},
  \citenamefont {Dhakal}, \citenamefont {Wang}, \citenamefont {Poudel},
  \citenamefont {Dimitri}, \citenamefont {Kabir}, \citenamefont {Sims},
  \citenamefont {Regmi}, \citenamefont {Gofryk}, \citenamefont {Kaczorowski},
  \citenamefont {Bansil},\ and\ \citenamefont {Neupane}}]{hosen_sras3}%
  \BibitemOpen
  \bibfield  {author} {\bibinfo {author} {\bibfnamefont {M.~Mofazzel}\
  \bibnamefont {Hosen}}, \bibinfo {author} {\bibfnamefont {Gyanendra}\
  \bibnamefont {Dhakal}}, \bibinfo {author} {\bibfnamefont {Baokai}\
  \bibnamefont {Wang}}, \bibinfo {author} {\bibfnamefont {Narayan}\
  \bibnamefont {Poudel}}, \bibinfo {author} {\bibfnamefont {Klauss}\
  \bibnamefont {Dimitri}}, \bibinfo {author} {\bibfnamefont {Firoza}\
  \bibnamefont {Kabir}}, \bibinfo {author} {\bibfnamefont {Christopher}\
  \bibnamefont {Sims}}, \bibinfo {author} {\bibfnamefont {Sabin}\ \bibnamefont
  {Regmi}}, \bibinfo {author} {\bibfnamefont {Krzysztof}\ \bibnamefont
  {Gofryk}}, \bibinfo {author} {\bibfnamefont {Dariusz}\ \bibnamefont
  {Kaczorowski}}, \bibinfo {author} {\bibfnamefont {Arun}\ \bibnamefont
  {Bansil}}, \ and\ \bibinfo {author} {\bibfnamefont {Madhab}\ \bibnamefont
  {Neupane}},\ }\bibfield  {title} {\enquote {\bibinfo {title} {Experimental
  observation of drumhead surface states in {SrAs}$_3$},}\ }\href {\doibase
  10.1038/s41598-020-59200-2} {\bibfield  {journal} {\bibinfo  {journal} {Sci.
  Rep.}\ }\textbf {\bibinfo {volume} {10}},\ \bibinfo {pages} {2776} (\bibinfo
  {year} {2020})}\BibitemShut {NoStop}%
\bibitem [{\citenamefont {Pezzini}\ \emph {et~al.}(2018)\citenamefont
  {Pezzini}, \citenamefont {Van~Delft}, \citenamefont {Schoop}, \citenamefont
  {Lotsch}, \citenamefont {Carrington}, \citenamefont {Katsnelson},
  \citenamefont {Hussey},\ and\ \citenamefont
  {Wiedmann}}]{pezzini_breakdownZrSiS}%
  \BibitemOpen
  \bibfield  {author} {\bibinfo {author} {\bibfnamefont {S.}~\bibnamefont
  {Pezzini}}, \bibinfo {author} {\bibfnamefont {M.R.}\ \bibnamefont
  {Van~Delft}}, \bibinfo {author} {\bibfnamefont {L.M.}\ \bibnamefont
  {Schoop}}, \bibinfo {author} {\bibfnamefont {B.V.}\ \bibnamefont {Lotsch}},
  \bibinfo {author} {\bibfnamefont {A.}~\bibnamefont {Carrington}}, \bibinfo
  {author} {\bibfnamefont {M.I.}\ \bibnamefont {Katsnelson}}, \bibinfo {author}
  {\bibfnamefont {N.E.}\ \bibnamefont {Hussey}}, \ and\ \bibinfo {author}
  {\bibfnamefont {S.}~\bibnamefont {Wiedmann}},\ }\bibfield  {title} {\enquote
  {\bibinfo {title} {Unconventional mass enhancement around the {Dirac} nodal
  loop in {ZrSiS}},}\ }\href {\doibase 10.1038/nphys4306} {\bibfield  {journal}
  {\bibinfo  {journal} {Nat. Phys.}\ }\textbf {\bibinfo {volume} {14}},\
  \bibinfo {pages} {178--183} (\bibinfo {year} {2018})}\BibitemShut {NoStop}%
\bibitem [{\citenamefont {Bradley}\ and\ \citenamefont
  {Davies}(1968)}]{bradley_magneticspacegroups}%
  \BibitemOpen
  \bibfield  {author} {\bibinfo {author} {\bibfnamefont {C.~J.}\ \bibnamefont
  {Bradley}}\ and\ \bibinfo {author} {\bibfnamefont {B.~L.}\ \bibnamefont
  {Davies}},\ }\bibfield  {title} {\enquote {\bibinfo {title} {Magnetic groups
  and their corepresentations},}\ }\href@noop {} {\bibfield  {journal}
  {\bibinfo  {journal} {Rev. Mod. Phys.}\ }\textbf {\bibinfo {volume} {40}},\
  \bibinfo {pages} {359--379} (\bibinfo {year} {1968})}\BibitemShut {NoStop}%
\bibitem [{\citenamefont {Yuan}\ \emph {et~al.}(2019)\citenamefont {Yuan},
  \citenamefont {Isobe},\ and\ \citenamefont {Fu}}]{yuannoah_highordervanhove}%
  \BibitemOpen
  \bibfield  {author} {\bibinfo {author} {\bibfnamefont {Noah F.~Q.}\
  \bibnamefont {Yuan}}, \bibinfo {author} {\bibfnamefont {Hiroki}\ \bibnamefont
  {Isobe}}, \ and\ \bibinfo {author} {\bibfnamefont {Liang}\ \bibnamefont
  {Fu}},\ }\bibfield  {title} {\enquote {\bibinfo {title} {Magic of high-order
  van {Hove} singularity},}\ }\href {\doibase 10.1038/s41467-019-13670-9}
  {\bibfield  {journal} {\bibinfo  {journal} {Nat. Commun.}\ }\textbf {\bibinfo
  {volume} {10}},\ \bibinfo {pages} {5769} (\bibinfo {year}
  {2019})}\BibitemShut {NoStop}%
\bibitem [{\citenamefont {Lifshitz}\ and\ \citenamefont
  {Kosevich}(1956)}]{lifshitz_kosevich_jetp}%
  \BibitemOpen
  \bibfield  {author} {\bibinfo {author} {\bibfnamefont {L.~M.}\ \bibnamefont
  {Lifshitz}}\ and\ \bibinfo {author} {\bibfnamefont {A.M.}\ \bibnamefont
  {Kosevich}},\ }\bibfield  {title} {\enquote {\bibinfo {title} {Theory of
  magnetic susceptibility in metals at low temperatures},}\ }\href@noop {}
  {\bibfield  {journal} {\bibinfo  {journal} {J. Exp. Theor. Phys.}\ }\textbf
  {\bibinfo {volume} {2}},\ \bibinfo {pages} {636} (\bibinfo {year}
  {1956})}\BibitemShut {NoStop}%
\bibitem [{\citenamefont {Falicov}\ and\ \citenamefont
  {Stachowiak}(1966)}]{falicov_dhva_magnesium}%
  \BibitemOpen
  \bibfield  {author} {\bibinfo {author} {\bibfnamefont {L.~M.}\ \bibnamefont
  {Falicov}}\ and\ \bibinfo {author} {\bibfnamefont {Henryk}\ \bibnamefont
  {Stachowiak}},\ }\bibfield  {title} {\enquote {\bibinfo {title} {Theory of
  the de {Haas}-van {Alphen} effect in a system of coupled orbits. application
  to magnesium},}\ }\href {\doibase 10.1103/PhysRev.147.505} {\bibfield
  {journal} {\bibinfo  {journal} {Phys. Rev.}\ }\textbf {\bibinfo {volume}
  {147}},\ \bibinfo {pages} {505--515} (\bibinfo {year} {1966})}\BibitemShut
  {NoStop}%
\bibitem [{\citenamefont {Zak}(1964)}]{zak_magnetictranslation}%
  \BibitemOpen
  \bibfield  {author} {\bibinfo {author} {\bibfnamefont {J.}~\bibnamefont
  {Zak}},\ }\bibfield  {title} {\enquote {\bibinfo {title} {Magnetic
  translation group},}\ }\href {\doibase 10.1103/PhysRev.134.A1602} {\bibfield
  {journal} {\bibinfo  {journal} {Phys. Rev.}\ }\textbf {\bibinfo {volume}
  {134}},\ \bibinfo {pages} {A1602--A1606} (\bibinfo {year}
  {1964})}\BibitemShut {NoStop}%
\bibitem [{\citenamefont {Brown}(1964)}]{Brown_magnetictranslation}%
  \BibitemOpen
  \bibfield  {author} {\bibinfo {author} {\bibfnamefont {E.}~\bibnamefont
  {Brown}},\ }\bibfield  {title} {\enquote {\bibinfo {title} {Bloch electrons
  in a uniform magnetic field},}\ }\href {\doibase 10.1103/PhysRev.133.A1038}
  {\bibfield  {journal} {\bibinfo  {journal} {Phys. Rev.}\ }\textbf {\bibinfo
  {volume} {133}},\ \bibinfo {pages} {A1038--A1044} (\bibinfo {year}
  {1964})}\BibitemShut {NoStop}%
\end{thebibliography}%


\begin{thebibliography}{14}%
\makeatletter
\providecommand \@ifxundefined [1]{%
 \@ifx{#1\undefined}
}%
\providecommand \@ifnum [1]{%
 \ifnum #1\expandafter \@firstoftwo
 \else \expandafter \@secondoftwo
 \fi
}%
\providecommand \@ifx [1]{%
 \ifx #1\expandafter \@firstoftwo
 \else \expandafter \@secondoftwo
 \fi
}%
\providecommand \natexlab [1]{#1}%
\providecommand \enquote  [1]{``#1''}%
\providecommand \bibnamefont  [1]{#1}%
\providecommand \bibfnamefont [1]{#1}%
\providecommand \citenamefont [1]{#1}%
\providecommand \href@noop [0]{\@secondoftwo}%
\providecommand \href [0]{\begingroup \@sanitize@url \@href}%
\providecommand \@href[1]{\@@startlink{#1}\@@href}%
\providecommand \@@href[1]{\endgroup#1\@@endlink}%
\providecommand \@sanitize@url [0]{\catcode `\\12\catcode `\$12\catcode
  `\&12\catcode `\#12\catcode `\^12\catcode `\_12\catcode `\%12\relax}%
\providecommand \@@startlink[1]{}%
\providecommand \@@endlink[0]{}%
\providecommand \url  [0]{\begingroup\@sanitize@url \@url }%
\providecommand \@url [1]{\endgroup\@href {#1}{\urlprefix }}%
\providecommand \urlprefix  [0]{URL }%
\providecommand \Eprint [0]{\href }%
\providecommand \doibase [0]{http://dx.doi.org/}%
\providecommand \selectlanguage [0]{\@gobble}%
\providecommand \bibinfo  [0]{\@secondoftwo}%
\providecommand \bibfield  [0]{\@secondoftwo}%
\providecommand \translation [1]{[#1]}%
\providecommand \BibitemOpen [0]{}%
\providecommand \bibitemStop [0]{}%
\providecommand \bibitemNoStop [0]{.\EOS\space}%
\providecommand \EOS [0]{\spacefactor3000\relax}%
\providecommand \BibitemShut  [1]{\csname bibitem#1\endcsname}%
\let\auto@bib@innerbib\@empty
\bibitem [{\citenamefont {Onsager}(1952)}]{onsager}%
  \BibitemOpen
  \bibfield  {author} {\bibinfo {author} {\bibfnamefont {L.}~\bibnamefont
  {Onsager}},\ }\bibfield  {title} {\enquote {\bibinfo {title} {Interpretation
  of the de {Haas}-van {Alphen} effect},}\ }\href {\doibase
  10.1080/14786440908521019} {\bibfield  {journal} {\bibinfo  {journal}
  {Philos. Mag.}\ }\textbf {\bibinfo {volume} {43}},\ \bibinfo {pages}
  {1006--1008} (\bibinfo {year} {1952})}\BibitemShut {NoStop}%
\bibitem [{\citenamefont {Lifshitz}\ and\ \citenamefont
  {Kosevich}(1956)}]{lifshitz_kosevich_jetp}%
  \BibitemOpen
  \bibfield  {author} {\bibinfo {author} {\bibfnamefont {L.~M.}\ \bibnamefont
  {Lifshitz}}\ and\ \bibinfo {author} {\bibfnamefont {A.M.}\ \bibnamefont
  {Kosevich}},\ }\bibfield  {title} {\enquote {\bibinfo {title} {Theory of
  magnetic susceptibility in metals at low temperatures},}\ }\href@noop {}
  {\bibfield  {journal} {\bibinfo  {journal} {J. Exp. Theor. Phys.}\ }\textbf
  {\bibinfo {volume} {2}},\ \bibinfo {pages} {636} (\bibinfo {year}
  {1956})}\BibitemShut {NoStop}%
\bibitem [{\citenamefont {Keller}(1958)}]{keller_correctedbohrsommerfeld}%
  \BibitemOpen
  \bibfield  {author} {\bibinfo {author} {\bibfnamefont {Joseph~B.}\
  \bibnamefont {Keller}},\ }\bibfield  {title} {\enquote {\bibinfo {title}
  {Corrected {Bohr-Sommerfeld} quantum conditions for nonseparable systems},}\
  }\href {\doibase http://dx.doi.org/10.1016/0003-4916(58)90032-0} {\bibfield
  {journal} {\bibinfo  {journal} {Ann. Phys. (N. Y.)}\ }\textbf {\bibinfo
  {volume} {4}},\ \bibinfo {pages} {180 -- 188} (\bibinfo {year}
  {1958})}\BibitemShut {NoStop}%
\bibitem [{\citenamefont {Roth}(1966)}]{rothII}%
  \BibitemOpen
  \bibfield  {author} {\bibinfo {author} {\bibfnamefont {Laura~M.}\
  \bibnamefont {Roth}},\ }\bibfield  {title} {\enquote {\bibinfo {title}
  {Semiclassical theory of magnetic energy levels and magnetic susceptibility
  of {Bloch} electrons},}\ }\href {\doibase 10.1103/PhysRev.145.434} {\bibfield
   {journal} {\bibinfo  {journal} {Phys. Rev.}\ }\textbf {\bibinfo {volume}
  {145}},\ \bibinfo {pages} {434--448} (\bibinfo {year} {1966})}\BibitemShut
  {NoStop}%
\bibitem [{\citenamefont {Mikitik}\ and\ \citenamefont
  {Sharlai}(1999)}]{mikitik_berryinmetal}%
  \BibitemOpen
  \bibfield  {author} {\bibinfo {author} {\bibfnamefont {G.~P.}\ \bibnamefont
  {Mikitik}}\ and\ \bibinfo {author} {\bibfnamefont {Yu.~V.}\ \bibnamefont
  {Sharlai}},\ }\bibfield  {title} {\enquote {\bibinfo {title} {Manifestation
  of {Berry}'s phase in metal physics},}\ }\href {\doibase
  10.1103/PhysRevLett.82.2147} {\bibfield  {journal} {\bibinfo  {journal}
  {Phys. Rev. Lett.}\ }\textbf {\bibinfo {volume} {82}},\ \bibinfo {pages}
  {2147--2150} (\bibinfo {year} {1999})}\BibitemShut {NoStop}%
\bibitem [{\citenamefont {Alexandradinata}\ and\ \citenamefont
  {Glazman}(2018)}]{100page}%
  \BibitemOpen
  \bibfield  {author} {\bibinfo {author} {\bibfnamefont {A.}~\bibnamefont
  {Alexandradinata}}\ and\ \bibinfo {author} {\bibfnamefont {Leonid}\
  \bibnamefont {Glazman}},\ }\bibfield  {title} {\enquote {\bibinfo {title}
  {Semiclassical theory of landau levels and magnetic breakdown in topological
  metals},}\ }\href {\doibase 10.1103/PhysRevB.97.144422} {\bibfield  {journal}
  {\bibinfo  {journal} {Phys. Rev. B}\ }\textbf {\bibinfo {volume} {97}},\
  \bibinfo {pages} {144422} (\bibinfo {year} {2018})}\BibitemShut {NoStop}%
\bibitem [{\citenamefont {Zener}\ and\ \citenamefont
  {Fowler}(1932)}]{zener_nonadiabaticcrossing}%
  \BibitemOpen
  \bibfield  {author} {\bibinfo {author} {\bibfnamefont {Clarence}\
  \bibnamefont {Zener}}\ and\ \bibinfo {author} {\bibfnamefont {Ralph~Howard}\
  \bibnamefont {Fowler}},\ }\bibfield  {title} {\enquote {\bibinfo {title}
  {Non-adiabatic crossing of energy levels},}\ }\href {\doibase
  10.1098/rspa.1932.0165} {\bibfield  {journal} {\bibinfo  {journal} {Proc. R.
  Soc. Lond. A}\ }\textbf {\bibinfo {volume} {137}},\ \bibinfo {pages}
  {696--702} (\bibinfo {year} {1932})}\BibitemShut {NoStop}%
\bibitem [{\citenamefont {Kaganov}\ and\ \citenamefont
  {Slutskin}(1983)}]{kaganov_coherentmagneticbreakdown}%
  \BibitemOpen
  \bibfield  {author} {\bibinfo {author} {\bibfnamefont {M.I.}\ \bibnamefont
  {Kaganov}}\ and\ \bibinfo {author} {\bibfnamefont {A.A.}\ \bibnamefont
  {Slutskin}},\ }\bibfield  {title} {\enquote {\bibinfo {title} {Coherent
  magnetic breakdown},}\ }\href {\doibase
  http://dx.doi.org/10.1016/0370-1573(83)90006-6} {\bibfield  {journal}
  {\bibinfo  {journal} {Phys. Rep.}\ }\textbf {\bibinfo {volume} {98}},\
  \bibinfo {pages} {189 -- 271} (\bibinfo {year} {1983})}\BibitemShut {NoStop}%
\bibitem [{\citenamefont {Falicov}\ and\ \citenamefont
  {Stachowiak}(1966)}]{falicov_dhva_magnesium}%
  \BibitemOpen
  \bibfield  {author} {\bibinfo {author} {\bibfnamefont {L.~M.}\ \bibnamefont
  {Falicov}}\ and\ \bibinfo {author} {\bibfnamefont {Henryk}\ \bibnamefont
  {Stachowiak}},\ }\bibfield  {title} {\enquote {\bibinfo {title} {Theory of
  the de {Haas}-van {Alphen} effect in a system of coupled orbits. application
  to magnesium},}\ }\href {\doibase 10.1103/PhysRev.147.505} {\bibfield
  {journal} {\bibinfo  {journal} {Phys. Rev.}\ }\textbf {\bibinfo {volume}
  {147}},\ \bibinfo {pages} {505--515} (\bibinfo {year} {1966})}\BibitemShut
  {NoStop}%
\bibitem [{\citenamefont {Xu}\ \emph {et~al.}(2017)\citenamefont {Xu},
  \citenamefont {Yu}, \citenamefont {Fang}, \citenamefont {Dai},\ and\
  \citenamefont {Weng}}]{qiunan_cap3}%
  \BibitemOpen
  \bibfield  {author} {\bibinfo {author} {\bibfnamefont {Qiunan}\ \bibnamefont
  {Xu}}, \bibinfo {author} {\bibfnamefont {Rui}\ \bibnamefont {Yu}}, \bibinfo
  {author} {\bibfnamefont {Zhong}\ \bibnamefont {Fang}}, \bibinfo {author}
  {\bibfnamefont {Xi}~\bibnamefont {Dai}}, \ and\ \bibinfo {author}
  {\bibfnamefont {Hongming}\ \bibnamefont {Weng}},\ }\bibfield  {title}
  {\enquote {\bibinfo {title} {Topological nodal line semimetals in the
  {CaP}$_{3}$ family of materials},}\ }\href {\doibase
  10.1103/PhysRevB.95.045136} {\bibfield  {journal} {\bibinfo  {journal} {Phys.
  Rev. B}\ }\textbf {\bibinfo {volume} {95}},\ \bibinfo {pages} {045136}
  (\bibinfo {year} {2017})}\BibitemShut {NoStop}%
\bibitem [{\citenamefont {Landau}\ and\ \citenamefont
  {Lifshitz}(2007)}]{landaulifshitz_quantummechanics}%
  \BibitemOpen
  \bibfield  {author} {\bibinfo {author} {\bibfnamefont {L.~D.}\ \bibnamefont
  {Landau}}\ and\ \bibinfo {author} {\bibfnamefont {E.~M.}\ \bibnamefont
  {Lifshitz}},\ }\href@noop {} {\emph {\bibinfo {title} {Quantum Mechanics}}}\
  (\bibinfo  {publisher} {Elsevier},\ \bibinfo {address} {Singapore},\ \bibinfo
  {year} {2007})\BibitemShut {NoStop}%
\bibitem [{\citenamefont {Zak}(1964)}]{zak_magnetictranslation}%
  \BibitemOpen
  \bibfield  {author} {\bibinfo {author} {\bibfnamefont {J.}~\bibnamefont
  {Zak}},\ }\bibfield  {title} {\enquote {\bibinfo {title} {Magnetic
  translation group},}\ }\href {\doibase 10.1103/PhysRev.134.A1602} {\bibfield
  {journal} {\bibinfo  {journal} {Phys. Rev.}\ }\textbf {\bibinfo {volume}
  {134}},\ \bibinfo {pages} {A1602--A1606} (\bibinfo {year}
  {1964})}\BibitemShut {NoStop}%
\bibitem [{\citenamefont {Brown}(1964)}]{Brown_magnetictranslation}%
  \BibitemOpen
  \bibfield  {author} {\bibinfo {author} {\bibfnamefont {E.}~\bibnamefont
  {Brown}},\ }\bibfield  {title} {\enquote {\bibinfo {title} {Bloch electrons
  in a uniform magnetic field},}\ }\href {\doibase 10.1103/PhysRev.133.A1038}
  {\bibfield  {journal} {\bibinfo  {journal} {Phys. Rev.}\ }\textbf {\bibinfo
  {volume} {133}},\ \bibinfo {pages} {A1038--A1044} (\bibinfo {year}
  {1964})}\BibitemShut {NoStop}%
\bibitem [{\citenamefont {Wang}\ \emph {et~al.}(2019)\citenamefont {Wang},
  \citenamefont {Duan}, \citenamefont {Glazman},\ and\ \citenamefont
  {Alexandradinata}}]{cwaa_landauquantization}%
  \BibitemOpen
  \bibfield  {author} {\bibinfo {author} {\bibfnamefont {Chong}\ \bibnamefont
  {Wang}}, \bibinfo {author} {\bibfnamefont {Wenhui}\ \bibnamefont {Duan}},
  \bibinfo {author} {\bibfnamefont {Leonid}\ \bibnamefont {Glazman}}, \ and\
  \bibinfo {author} {\bibfnamefont {A.}~\bibnamefont {Alexandradinata}},\
  }\bibfield  {title} {\enquote {\bibinfo {title} {Landau quantization of
  nearly degenerate bands and full symmetry classification of landau level
  crossings},}\ }\href {\doibase 10.1103/PhysRevB.100.014442} {\bibfield
  {journal} {\bibinfo  {journal} {Phys. Rev. B}\ }\textbf {\bibinfo {volume}
  {100}},\ \bibinfo {pages} {014442} (\bibinfo {year} {2019})}\BibitemShut
  {NoStop}%
\end{thebibliography}%

\clearpage

\end{document}


\title{Supplemental material for `Diabolical touching point in the magnetic energy levels of topological nodal-line metals'}

\author{Chong Wang}\affiliation{Department of Physics, Carnegie Mellon University, Pittsburgh, Pennsylvania 15213, USA}
\author{Zhongyi Zhang}\affiliation{Institute of Physics, Chinese Academy of Sciences, Beijing 100080, China}
\author{Chen Fang}\affiliation{Institute of Physics, Chinese Academy of Sciences, Beijing 100080, China}
\author{A. Alexandradinata}\affiliation{Department of Physics and Institute for Condensed Matter Theory, University of Illinois at Urbana-Champaign, Urbana, Illinois 61801, USA}

\maketitle

{\tableofcontents \par}

\section{Magnetic energy levels of two-pocket model}

For our two-pocket model of a nodal-line metal subject to magnetic breakdown, we offer a more detailed description of the quantization rule [Eq.\ (4) in main text]  and sum-over-histories formula [Eq.\ (6) in main text] for the density of states (DOS). 

\subsection{Quantization rule}

As described in the main text, at fixed wavenumber $k_z$ and fixed energy $\varepsilon$, the model band structure is mapped onto an oriented graph composed of two two-in-two-out vertices and four edges (labelled by $\alpha=1\ldots 4$). The two vertices are physically associated to focal points of magnetic breakdown. Each edge is oriented according to the Lorentz force under a field in the $-z$ direction. The orientation allows  to uniquely define a start and end point for edges -- each edge starts at a vertex and ends at a distinct vertex. We will present a quantization rule for magnetic energy levels that is valid in the regime $S_\text{min}/B\gg 1$, where $S_\text{min}$ as the area of the smallest loop in the graph (corresponding to one of two sausages).

Let us define $U_{\beta \alpha}$ as the complex-valued amplitude for an electron to (i) traverse said edge $\alpha$ and accumulate a phase factor $\exp (i\varphi_\alpha)$ (to be specified below), and (ii) to subsequently scatter (at a vertex at the end point of $\alpha$) into a possibly-distinct edge $\beta$ with amplitude $V_{\beta \alpha}$. Collecting all 16 elements into a $4\times 4$ matrix, $U$ can be expressed as the product of $V$ with a diagonal matrix:
\begin{eqnarray}
  U(\var,k_z,B) & = & \left( \begin{array}{cccc}
    0 & \tau \mathe^{\mathi \omega} & - \rho & 0\\
    \tau \mathe^{\mathi \omega} & 0 & 0 & - \rho\\
    \rho & 0 & 0 & \tau \mathe^{- \mathi \omega}\\
    0 & \rho & \tau \mathe^{- \mathi \omega} & 0
  \end{array} \right) \left( \begin{array}{cccc}
    \mathe^{\mathi \varphi_1} & 0 & 0 & 0\\
    0 & \mathe^{\mathi \varphi_2} & 0 & 0\\
    0 & 0 & \mathe^{\mathi \varphi_3} & 0\\
    0 & 0 & 0 & \mathe^{\mathi \varphi_4}
  \end{array} \right).\la{dynamicalminimal}
\end{eqnarray}
Let us first describe the diagonal matrix which encodes process (i). The phase accumulated on each edge $\alpha$  is 
\e{ \varphi_\alpha=-B^{-1}\int_\alpha k_x^\alpha \mathd k_y + m_{\alpha}\f{\pi}{2}+ \varphi^g_{\alpha}. \la{definevarphi}}
The first term is the leading-order dynamical phase~\cite{onsager,lifshitz_kosevich_jetp},
with $k_x=k_x^{\alpha}(k_y,k_z,\varepsilon)$ defining a $\bk$-curve corresponding to the edge $\alpha$, and the above line integral is oriented by the Lorentz force. $k_x^{2,3}$ are single-valued functions, however
$k_x^{1}$ (and also $k_x^{4}$)  is a multivalued function with three branches because edge $1$ (resp.\ edge $4$) contains two {turning points}, where $\partial k^{\alpha}_x/\partial k_y=0$. The subleading correction to $\varphi_\alpha$ sums the Maslov phase~\cite{keller_correctedbohrsommerfeld} ($\pi/2$ times the number of turning points) and the geometric Berry phase $\varphi^g_{\alpha}$~\cite{rothII,mikitik_berryinmetal}.  

Each two-in-two-out vertex is associated to four nonzero amplitudes $V_{\beta \alpha}$ which can be collected into a two-by-two, unitary `scattering matrix'~\cite{100page}:
\e{\mathcal{S} (\mu)  =  \matrixtwo{
    \tau \mathe^{\mathi \omega}}{ - \rho}{\rho}{\tau \mathe^{- \mathi \omega}}\la{scatmatrix}}
with $\mu = v_z^2 k_z^2 / 2 v_x v_y B$, $\rho = \exp (- \mathpi \mu)$,  $\tau = \sqrt{1 - \rho^2}$ and $\omega = \mu - \mu \ln \mu + \arg [\Gamma (\mathi \mu)] + \mathpi / 4$.  $|\mathcal{S}_{12}|^2=\exp(-2\pi \mu)$ is identified with the Landau-Zener tunneling probability~\cite{zener_nonadiabaticcrossing}.  For a pair of  edges $(\beta,\alpha)$ that are not connected by a vertex, we define $V_{\beta \alpha}=0.$

The quantization rule determines energy levels  by the condition that the four-component wave function (one component for each edge) is single-valued over the graph. This rule is conveniently expressed as~\cite{kaganov_coherentmagneticbreakdown,100page}
\begin{eqnarray}
  \det [I - U(E,k_z,B)] & = & 0, \la{determinant}
\end{eqnarray}
where $I$ is the $4\times 4$ identity matrix. Inserting our expression for $U$ into \q{determinant} and evaluating the determinant, one obtains the quantization rule for our two-pocket model, as expressed in Eq.\ (4) of the main text.
\subsection{Density of states}

The density of states (DOS)  involves an interference of  all possible Feynman loops, even loops that are integer repetitions of a fundamental loop~\cite{falicov_dhva_magnesium}. Let us therefore construct a four-by-four matrix $P$ from the geometric series:
\begin{eqnarray}
  P & = & \sum_{n = 1}^{\infty} U^n, \as   P = UP + U, \la{recursive}
\end{eqnarray}
such that $P_{\ab}$ is a sum of amplitudes for trajectories of all possible lengths indexed by $n$, where the
length of a path is the number of edges traversed. The second equality in \q{recursive} can be viewed element by element as an inhomogeneous system of linear equations for sixteen variables $P_{\ab}$; $P=U(1-U)^{-1}$ is easily calculated   once $U$ is determined from \q{dynamicalminimal}.

The $P$ matrix is related to the DOS $\nu$ as
\begin{eqnarray}
  \nu (\varepsilon) & \approx & \frac{\mathcal{D}}{2 \mathpi} \left|
  \sum_{\alpha} (\partial_{\varepsilon} \varphi_{\alpha}) [2 \tmop{Re}
  (P_{\alpha \alpha}) + 1] \right|_{\tmmathbf{\varphi} \rightarrow
  \tmmathbf{\varphi}+ \mathi \tmmathbf{0}^+} \label{rho-to-P}
\end{eqnarray}
with a correction of relative magnitude $|dS_\square/d E|/|dS_\text{min}/d E|$; we remind the reader that $S_\square$ is the area of the $\bk$-rectangle illustrated in the inset of Fig.\ 2(a) of the main text.  $\mathcal{D}$ above is the degeneracy of a single Landau level. The right-hand side of Eq. (\ref{rho-to-P}) depends on the energy $\varepsilon$ through
${\varphi}_{\alpha}$ and the scattering matrix. To make the right-hand side of Eq. (\ref{rho-to-P}) well-defined as a generalized function, we replace ${\varphi}_\alpha \rightarrow {\varphi}_\alpha+ \mathi
0^+$ for all $\alpha$, with $0^+$ a positive infinitesimal.

In comparison with existing literature, it was proposed by Kaganov and Slutskin [in \ocite{kaganov_coherentmagneticbreakdown}]
that the density of states  equals 
\begin{eqnarray}
\nu (\varepsilon) & \stackrel{?}{=} & \frac{\mathcal{D}}{2 \mathpi} 
  \sum_{\alpha} \big|(\partial_{\varepsilon} \varphi_{\alpha})\big| [2 \tmop{Re}
  (P_{\alpha \alpha}) + 1]_{\tmmathbf{\varphi} \rightarrow
  \tmmathbf{\varphi}+ \mathi \tmmathbf{0}^+};  \label{rhokaganov}
\end{eqnarray}
$|\partial_{\varepsilon} \varphi_{\alpha}|$, to leading order in $B^{\text{-}\sma{1}}$, equals the time $T_{\alpha}$ taken for an electron to traverse edge $\alpha$ following the semiclassical equation of motion. We observe that \q{rhokaganov}  differs from our formula [cf.\ \q{rho-to-P}] in where the  absolute value symbol is placed, which affects whether certain amplitudes add constructively or subtract destructively. 

\begin{figure}[H]
  \centering
  \includegraphics[width=0.8\textwidth]{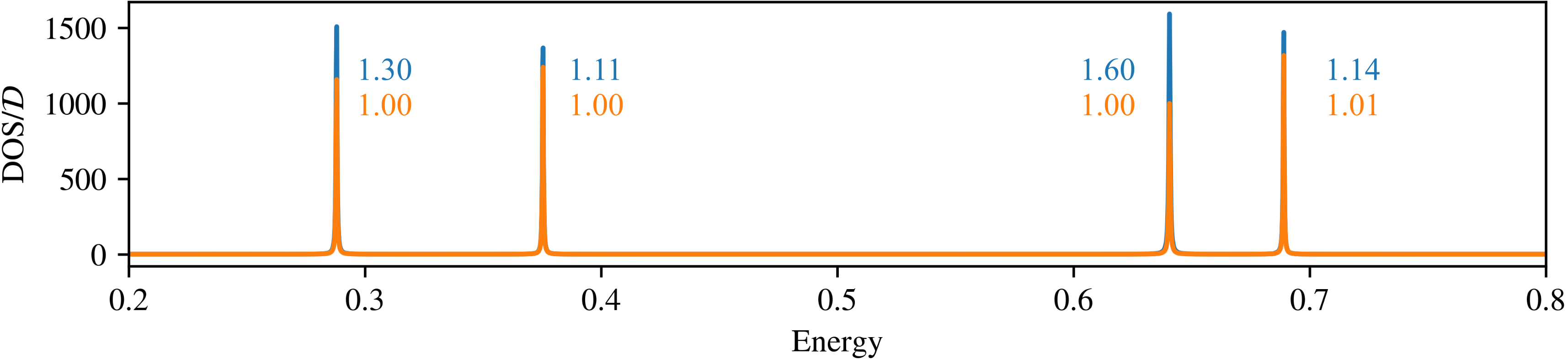}
  \caption{Density of states calculated by Eq. (\ref{rhokaganov}) (resp. blue) and Eq. (\ref{rho-to-P}) (resp. yellow) at $k_z = 2.5$ and $1/B = 3.0$. The area under the blue (resp. yellow) line for each peak is annotated beside the peak with blue (resp. yellow) text. The model parameters have been specified in the main text. \label{fig:compare_dos}}
\end{figure}

Postponing a detailed, analytic proof of our formula to a follow-up publication, we offer here a simple numerical demonstration that the Kaganov-Slutksin formula is incorrect.
Figure \ref{fig:compare_dos} shows the density of states (normalized by $\mathcal{D}$) as a function of energy for several Landau levels. Each Landau level corresponds to a delta-function peak that has been broadened into a Lorentzian-like wave form by a finite lifetime. The specific regularization we choose is $\varphi_{\alpha} \ri \varphi_{\alpha} + \mathi 0.00002 \varepsilon_0 T_\alpha$. Being a regularization of a delta function, the integration of the density of states  over this Lorentzian should give one (in units of $\mathcal{D}$). We find numerically that our formula [Eq. (\ref{rho-to-P})] correctly yields $1.0$ while the analogous integral for the Kaganov-Slutskin formula [Eq.\ (\ref{rhokaganov})] significantly deviates from $1.0$. 

\section{Additional analysis of CaP$_3$}

We derive in \s{sec:effmassmodel} the effective-mass model that was used in the main text. \s{sec:landaucap3} provides more details on the calculation of Landau levels.

Throughout this section, we use \AA~as the length unit and eV as the energy unit. We also neglect the spin-orbit interaction, whose energy scale is predicted to be in the range $4$ to $30$ meV~\cite{qiunan_cap3}. This energy scale is comparable to the Landau-level spacing (presented below) at fields between $5$ T and $10$ T, therefore our numerical simulations should not be viewed as a quantitative model of the magnetic energy levels of  CaP$_3$; rather, they should be viewed as proof of principle for the existence of Landau-Dirac points in CaP$_3$. (We remind the reader that Landau-Dirac points are perturbatively robust to spatial-inversion-symmetric perturbations -- including the spin-orbit interaction.) A quantitative model of magnetized CaP$_3$ must account for the Zeeman interaction and a realistic spin-orbit interaction -- a project we leave to future investigations. 

\subsection{Effective-mass model of CaP$_3$}\la{sec:effmassmodel}

The effective-mass model of CaP$_3$ around the Y point can be found in Ref.~[\onlinecite{qiunan_cap3}]. The spinless Hamiltonian is written as 
\begin{equation}
  \begin{split}
  &H(\bm{k})=\sum_{i=0}^3 g_i \sigma_i\\
     &g_0=a_0+a_1k_a^2+a_2k_b^2+a_3k_c^2,\\
     &g_1=0,\\
     &g_2=\alpha k_a+\beta k_b+\gamma k_c,\\
     &g_3=m_0+m_1k_a^2+m_2k_b^2+m_3k_c^2,\\
     &a_0=-0.091,a_1=1.671,a_2=14.372,a_3=2.394,\\
     &m_0=-0.142,m_1=10.438,m_2=19.138,m_3=11.91,\\
     &\alpha =1.773,\beta =0.001,\gamma =-2.096.\\
  \end{split}
\end{equation}
Here, $\sigma_{1,2,3}$ are Pauli matrices, $\sigma_0$ is the two-by-two identity matrix, $k_a$, $k_b$ and $k_c$ are reduced coordinates corresponding to the reciprocal-lattice vector of CaP$_3$. Using Cartesian coordinates, the parameters $g_i$ becomes:
\begin{equation}
  \begin{split}
    &g_0=-0.091+5.348k_x^2-2.91k_xk_y+49.303k_y^2-1.136k_xk_z-31.67k_yk_z+13.83k_z^2,\\
    &g_2=3.172 k_x-0.8611 k_y-4.297k_z,\\
    &g_3=33.40 k_x^2-18.18k_xk_y-7.098k_xk_z+67.6 k_y^2-40.65 k_yk_z+49.83 k_z^2-0.142.
  \end{split}
\end{equation}

To simplify the Hamiltonian, we perform a series of transformations:

\begin{itemize}
    \item We rotate the system such that $g_2(\bm{k})$ only depends on $k_z$:
        \begin{equation}
          (k_x,k_y,k_z)=(k_x^\prime,k_y^\prime,k_z^\prime)\left(
        \begin{array}{ccc}
         -0.7665& 0.2081& -0.6076\\
         0.262& 0.9651& 0 \\
         0.5863& -0.159& -0.7943 \\
        \end{array}
        \right).
        \end{equation}
        
    \item We perform a unitary transformation with $U=(1+i\sigma_z)/\sqrt{2}$ such that $\sigma_x\rightarrow-\sigma_y,\sigma_y\rightarrow\sigma_x$.
    
    \item We remove the $k_x k_y$ term in $g_0$ by defining
    \begin{equation}
  \left(
\begin{array}{c}
 k_x^{\prime\prime} \\
 k_y^{\prime\prime} \\
\end{array}
\right)=\left(
\begin{array}{cc}
 0.4295& 0.90305\\
 -0.90306& 0.4295\\
\end{array}
\right) \left(
\begin{array}{c}
 k_x^{\prime} \\
 k_y^{\prime} \\
\end{array}
\right),\quad k_z^{\prime\prime}=k_z^{\prime}.
\end{equation}

    \item We shift the energy $g_0\rightarrow g_0+\mu_0=g_0+0.01741$ such that the hole and electron pockets enclose the same volume at zero energy.
    
\end{itemize}

After all the above transformations, we obtain
\begin{equation}
  \begin{split}
    &g_0=54.68 k_x^{\prime\prime2}+11.33 k_x^{\prime\prime} k_z^{\prime\prime}+5.188  k_y^{\prime\prime2}-1.75k_y^{\prime\prime} k_z^{\prime\prime}+8.608k_z^{\prime\prime2}-0.091,\\
    &g_1=5.409 k_z^{\prime\prime},\\
    &g_3=77.73 k_x^{\prime\prime2}-20.66k_x^{\prime\prime} k_y^{\prime\prime}+15.74 k_x^{\prime\prime} k_z^{\prime\prime}+28.6  k_y^{\prime\prime2}-3.75  k_y^{\prime\prime} k_z^{\prime\prime}+44.49  k_z^{\prime\prime2}-0.142.
  \end{split}\label{effmass_cap}
\end{equation}
These coefficients of the transformed Hamiltonian $H(\bm{k}^{\prime\prime})=\sum_{i=0}^3 g_i \sigma_i$ will be used in the subsequent  calculation of Landau levels. 
The convenience attained with this coordinate system is that  $g_1$ vanishes within the $(k_z''=0)$ plane, so that the two-by-two Hamiltonian is diagonal with diagonal elements $H_{11}$ and $H_{22}$; we thus  refer to these coordinates as the \textit{diagonal coordinate system}, and  henceforth  drop the primes on $k_x$, $k_y$ and $k_z$.


It is convenient to remove the $k_x k_y$ term in $H_{11}$ by
\begin{equation}
  \left(
\begin{array}{c}
 k_x^{\prime} \\
 k_y^{\prime} \\
\end{array}
\right)=\left(
\begin{array}{cc}
 -0.9947 & 0.103 \\
 -0.103 & -0.9947\\
\end{array}
\right) \left(
\begin{array}{c}
 k_x \\
 k_y \\
\end{array}
\right),
\end{equation}
and the $k_xk_y$ term in $H_{22}$ by
\begin{equation}
  \left(
\begin{array}{c}
 k_x^{\prime\prime} \\
 k_y^{\prime\prime} \\
\end{array}
\right)=\left(
\begin{array}{cc}
 -0.7&0.7133\\
 -0.7133&-0.7 \\
\end{array}
\right) \left(
\begin{array}{c}
 k_x \\
 k_y \\
\end{array}
\right),
\end{equation}
giving finally $H_{11}=-0.2156 +133.485 k_x^{\prime2} + 32.7191 k_y^{\prime2}$, $H_{22}=0.06841-33.56 k_x^{\prime\prime2} -12.9 k_y^{\prime\prime2}$. 

\subsection{Landau levels of CaP$_3$} \la{sec:landaucap3}

\begin{figure}
  \begin{minipage}[H]{0.5\linewidth}
\centering
\includegraphics[width=3in]{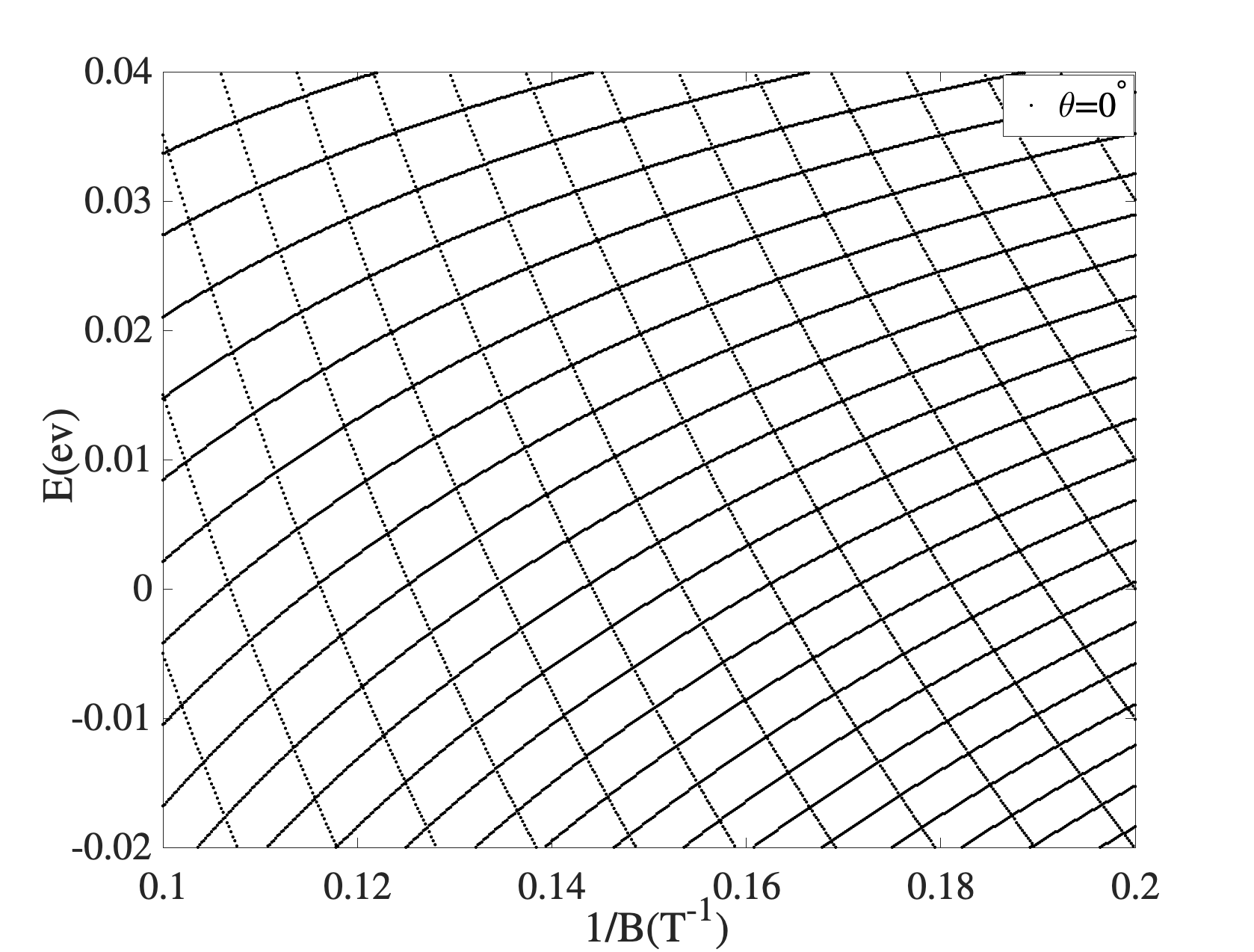}
\label{fig:side:a}
\end{minipage}%
\begin{minipage}[H]{0.5\linewidth}
\centering
\includegraphics[width=3in]{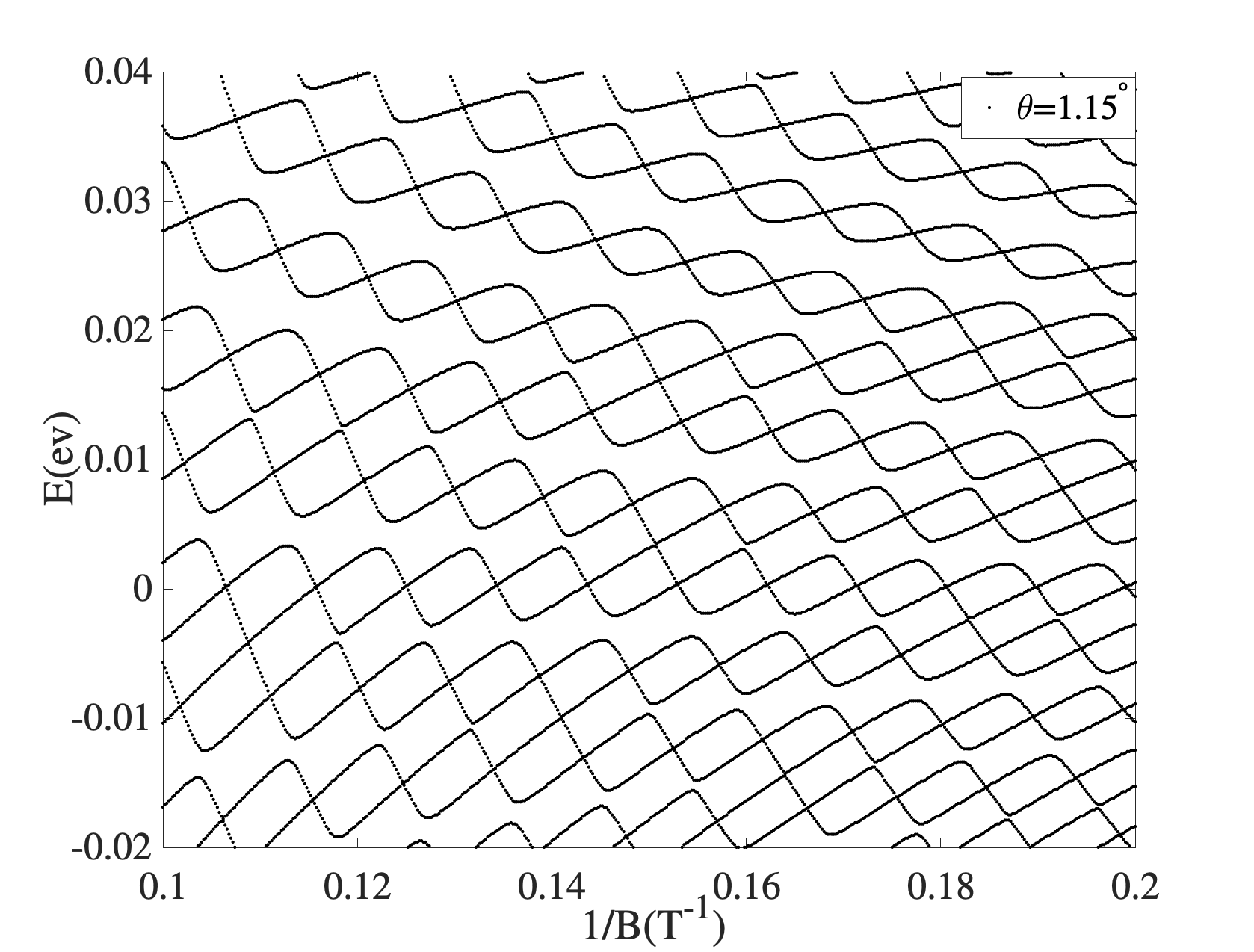}
\end{minipage}
\caption{Landau levels in the $k_z=0$ plane (field-fixed coordinates) for two field orientations.\label{kz0ll}}
\end{figure}

We have performed calculations for various orientations of the magnetic field, relative to a fixed crystallographic axis. For each orientation, we always adopt a right-handed Euclidean coordinate system such that the magnetic field lies in the $-z$ direction; such a system is uniquely defined up to rotations about the $z$ axis, and will be referred to as \textit{field-fixed coordinates}. No matter the field orientation, every $\bm{k}$ point in the $k_z=0$ plane (in field-fixed coordinates)  is mapped to itself by spatial-inversion symmetry.


Generally, the field-fixed coordinates differ from the diagonal coordinates defined in \s{sec:effmassmodel}.
When both coordinate systems coincide, the magnetic energy levels (in the $k_z=0$ plane) are obtained from separately diagonalizing the Peierls-Onsager Hamiltonians $H_{11}(\bK)$ and $H_{22}(\bK)$. These are obtained from $H_{11}(\bk)$ and $H_{22}(\bk)$ (defined in the previous subsection) by the standard Peierls substitution: $(k_x,k_y)\ri (k_x+By,k_y)$, in the Landau electromagnetic gauge $\bA=(By,0,0)$, with $[y,k_y]=i$. (Note that the principal coordinate axes for $H_{11}(\bk)$ and $H_{22}(\bk)$ are distinct, so the above Peierls substitutions are really carried out in different coordinate systems; this subtlety will affect neither the energy levels nor our conclusions about symmetry-protected crossings.) For either Hamiltonian, the Landau-level  wave functions are labelled by $n\in 0,1,2\ldots$ and wavenumber $k_x$, and have the analytic form: $\psi_{nk_x}=\exp (\mathi k_x x) \text{H}_n(\alpha y)$, where $\alpha$ is a constant and $\text{H}_n$ is the Hermite polynomial~\cite{landaulifshitz_quantummechanics}. The corresponding magnetic energy levels are presented in  Fig.~\ref{kz0ll}(a).

The following symmetry analysis is useful to determine the stability of the Landau-Dirac crossings. The inversion symmetry acts on eigenstates $\{\psi_{n k_x}\}_{n,k_x}$ of $H_{11}(\bK)$ as $\hat{\mathfrak{i}} \psi_{n k_x}(y) = (-1)^n \psi_{n ;-k_x}(y)$, where $(-1)^n$ originates from inverting the Hermite function. It is convenient to map $-k_x$ back to $k_x$ by the magnetic translation:
\begin{equation}
    \hat{t}(\bm{R}) = \mathe^{- \mathi [\hat{\tmmathbf{p}} +\tmmathbf{A} (\tmmathbf{r}) -\tmmathbf{B}
\times \tmmathbf{r}] \cdot \tmmathbf{R}}
\end{equation}
which is also a symmetry of the Peierls-Onsager Hamiltonian~\cite{zak_magnetictranslation,Brown_magnetictranslation}. In particular, $\hat{t}(2 k_x \hat{\bm{y}}/B) \psi_{n -k_x}(y) = \psi_{n k_x}(y)$, with $\hat{\bm{y}}$ the unit vector in the $y$ direction. Therefore, we deduce
\begin{equation}
\hat{\mathfrak{i}} \hat{t}(2 k_x \hat{\bm{y}}/B) \psi_{n k_x}(y) = (-1)^n \psi_{nk_x}(y),
\end{equation}
which states that adjacent Landau levels [for $H_{11}(\bK)$]  belong to opposite eigenspaces of the operator $\hat{\mathfrak{i}} \hat{t}(2 k_x \hat{\bm{y}}/B)$. 
Based on this symmetry analysis, and the analogous analysis for  $H_{22}(\bK)$, we demonstrated in the main text that half the Landau-Dirac crossings in Fig.~\ref{kz0ll}(a) are crossings between states in different $\hat{\mathfrak{i}} \hat{t}(2 k_x \hat{\bm{y}}/B)$ representations; the other half are protected by a different symmetry which exists only when the field-fixed and diagonal coordinates coincide. 

Thus one expects that half the Landau-Dirac crossings in  Fig.~\ref{kz0ll}(a) destabilize upon tilting of the field. This is confirmed by a calculation for which the field is tilted in the {$x-z$} plane (diagonal coordinates) by an angle $\theta = 1.15^\circ$ relative to the $-z$ axis (diagonal coordinates);
the resultant Landau levels in the $k_z=0$ plane (field-fixed coordinates) are shown in Fig. \ref{kz0ll}(b). For this calculation, analytic solutions are not available and thus numerical diagonalization is performed by standard techniques~\cite{cwaa_landauquantization}. 

\begin{figure}
\centering
\includegraphics[width=12cm]{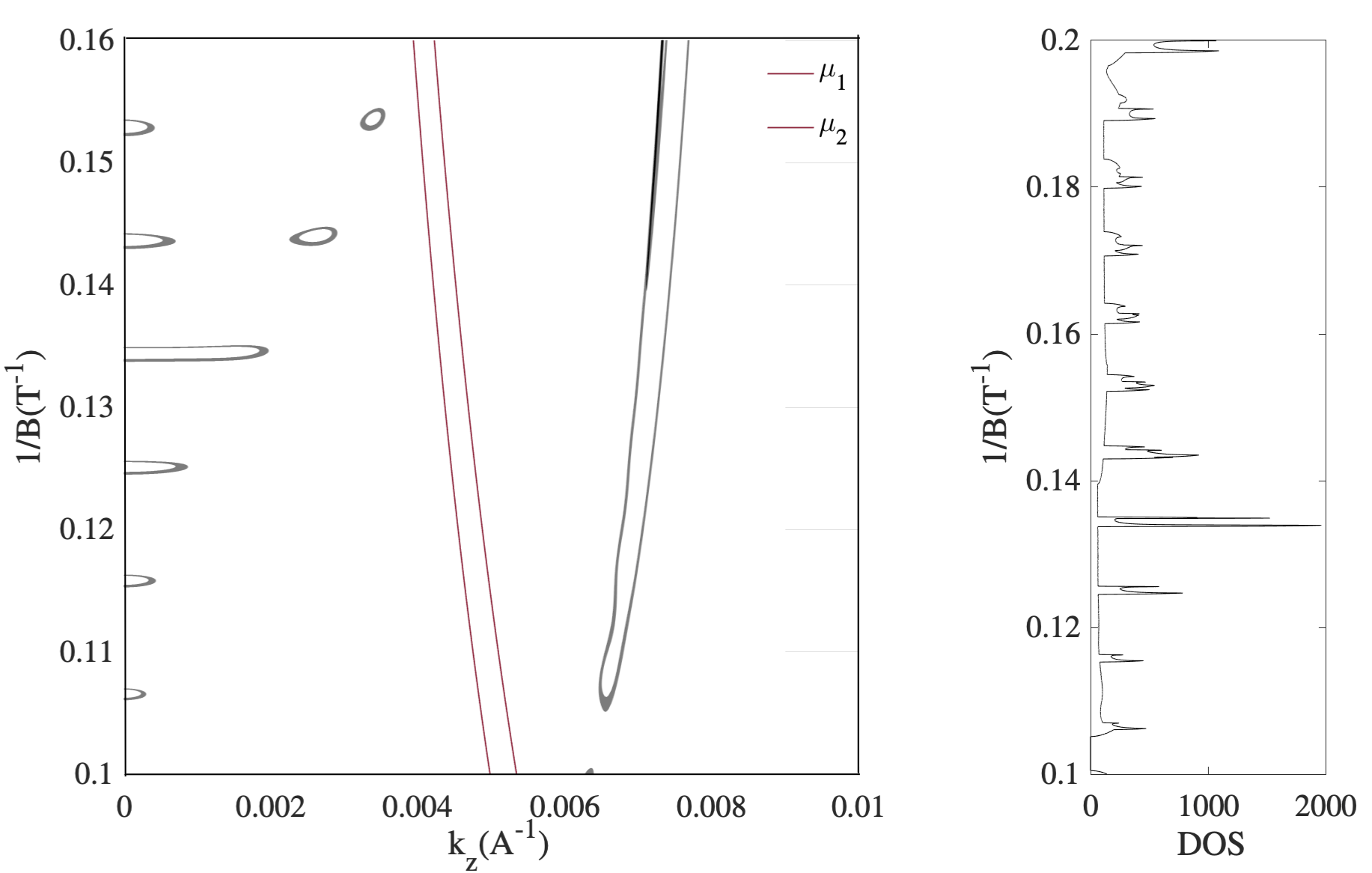}
\caption{For the spinless model of CaP$_3$, we plot the  Landau-Fermi surface over (1/B,$k_z$)-space in panel (a), and the corresponding density of states in panel (b).  \label{cap3ld}}
\end{figure}

Finally, we show in Fig.~\ref{cap3ld}(a) the Landau-Fermi surfaces of the magnetic energy levels at zero energy (the charge-neutral point); this plot is an expanded version of Fig.~3(b) in the main text. The type-I Landau-Fermi surfaces result in batman peaks in the density of states, as illustrated in the right panel. 

Observe that the type-I Landau-Fermi surfaces are confined (roughly) to the left half of the plot, where quantum tunneling is non-negligible. To make this observation precise, we calculate the Landau-Zener parameter $\mu(B^{\text{-}\sma{1}},k_z)$ for each of the four breakdown regions in the graph of CaP$_3$. $\mu$ is calculated by linearizing the effective-mass Hamiltonian at each of the four sausage links, then computing $\mu=S_\square/8B$ (with $S_\square$ the area of the $\bk$-rectangle inscribed by the hyperbolic band contours), just as we did for our two-pocket model in the main text. Accounting for $\inv$ symmetry, there are two (instead of four) independent values of $\mu$, given by $\mu_1{=} 5.571\times10^{4}k_z^2/B$  and $\mu_2{=}6.4\times10^{4}k_z^2/B$.   $\mu_1{=}\mu_2{=}1/\pi$ defines two curves in $(B^{\text{-}\sma{1}},k_z)$-space colored red in Fig.~\ref{cap3ld}(a). To the right of both curves, the tunneling probability $P_{1,2}=e^{-2\pi \mu_{1,2}}< e^{-2}$ is negligible, and the Landau-Fermi surface is determined by four independent cyclotron orbits over the four pockets. 

\bibliography{bib_Apr2018}